\documentclass[aps,prb,superscriptaddress,amsmath,amssymb,floats,twocolumn,longbibliography,floatfix]{revtex4-1}
\pdfoutput=1
\usepackage{graphicx}
\usepackage{hyperref}
\renewcommand{\paragraph}{\textit}

\usepackage{color}

\begin{document}

\title{Magnetic ground state and magnon-phonon interaction in multiferroic \texorpdfstring{h-YMnO$_3$}{h-YMnO3}}

\author{S. L. Holm}
\email[]{sonja@fys.ku.dk}
\affiliation{Nanoscience Center, Niels Bohr Institute, University of Copenhagen, 2100 Copenhagen {\O}, Denmark}

\author{A. Kreisel}
\affiliation{Niels Bohr Institute, Juliane Maries Vej 30, University of Copenhagen, 2100 Copenhagen {\O}, Denmark}
\affiliation{Institut f\" ur Theoretische Physik, Universit\" at Leipzig, D-04103 Leipzig, Germany}

\author{T. K. Sch\"affer}
\author{A. Bakke}
\author{M. Bertelsen}
\author{U. B. Hansen}
\affiliation{Nanoscience Center, Niels Bohr Institute, University of Copenhagen, 2100 Copenhagen {\O}, Denmark}

\author{M. Retuerto}
\affiliation{Nanoscience Center, Niels Bohr Institute, University of Copenhagen, 2100 Copenhagen {\O}, Denmark}
\affiliation{Instituto de Catálisis y Petroleoquímica, Consejo Superior de Investigaciones Cientificas, Cantoblanco E-28049, Madrid, Spain}

\author{J. Larsen}
\affiliation{Department of Physics, Technical University of Denmark, 2800 Kongens Lyngby, Denmark}

\author{D. Prabhakaran}
\affiliation{Clarendon Laboratory, Department of Physics, Oxford University, Parks Road, Oxford, OX1 3PU, United Kingdom}

\author{P. P. Deen}
\affiliation{European Spallation Source ERIC, 22363 Lund, Sweden}

\author{Z. Yamani}
\affiliation{Canadian Neutron Beam Centre, Chalk River Laboratories, Ontario, Canada}

\author{J. O. Birk}
\affiliation{Nanoscience Center, Niels Bohr Institute, University of Copenhagen, 2100 Copenhagen {\O}, Denmark}
\affiliation{Laboratory for Neutron Scattering and Imaging, Paul Scherrer Institute, 5232 Villigen PSI, Switzerland}

\author{U. Stuhr}
\author{Ch. Niedermayer}
\author{A. L. Fennell}
\affiliation{Laboratory for Neutron Scattering and Imaging, Paul Scherrer Institute, 5232 Villigen PSI, Switzerland}

\author{B. M. Andersen}
\author{K. Lefmann}
\affiliation{Niels Bohr Institute, Juliane Maries Vej 30, University of Copenhagen, 2100 Copenhagen {\O}, Denmark}

\date{\today}

\begin{abstract}
Inelastic neutron scattering has been used to study the magneto-elastic excitations in the multiferroic manganite hexagonal YMnO$_3$. An avoided crossing is found between magnon and phonon modes close to the Brillouin zone boundary in the $(a,b)$-plane. Neutron polarization analysis reveals that this mode has mixed magnon-phonon character. An external magnetic field along the $c$-axis is observed to cause a linear field-induced splitting of one of the spin wave branches. A theoretical description is performed, using a Heisenberg model of localized spins, acoustic phonon modes and a magneto-elastic coupling via the single-ion magnetostriction. The model quantitatively reproduces the dispersion and intensities of all modes in the full Brillouin zone, describes the observed magnon-phonon hybridized modes, and quantifies the magneto-elastic coupling. 
The combined information, including the field-induced magnon splitting, allows us to exclude several of the earlier proposed models and point to the correct magnetic ground state %of the $P6_3cm$ 
symmetry, and provides an effective dynamic model relevant for the multiferroic hexagonal manganites.
\end{abstract}


\maketitle

\section{Introduction}
Multiferroic materials display an intriguing coupling between structural, magnetic and electronic order. 
These properties make the material group interesting for applications in multi-functional devices, \textit{e.g.}\ as transducers, actuators, %capacitors, spintronics, 
or multi-memory devices \cite{Hill2000,Lee2008a,Catalan2009}. 
Since most known multiferroics are functional only at low temperatures, however, the route to practical application goes through an improved understanding of their basic material properties \cite{Spalding2005,Cheong2007}.

To determine the mechanisms behind multiferroicity, the magnetic and structural dynamics of the materials are studied, using \textit{e.g.} Raman or THz spectroscopy, or inelastic neutron scattering (INS). In type-II multiferroics, where the ferroelectric ordering generally takes place at the same temperature as the (antiferro-) magnetic ordering, these techniques have revealed a hybridization of magnons and electrically active optical phonons, known as electromagnons \cite{senff2007}. 
In type-I multiferroics, where the ferroelectric transition takes place at higher temperatures than the magnetic ordering, INS was used to measure magnon dispersions, obtaining the spin-spin interactions, \cite{vajk05,jeong12,matsuda12,jeong14}. The spin-lattice coupling, involved in multiferroicity, has been studied in both CuCrO$_2$ \cite{park2016} and in the only room temperature multiferroic, BiFeO$_3$ \cite{schneeloch2015}.

\begin{figure}[tb]
\includegraphics[width=\linewidth]{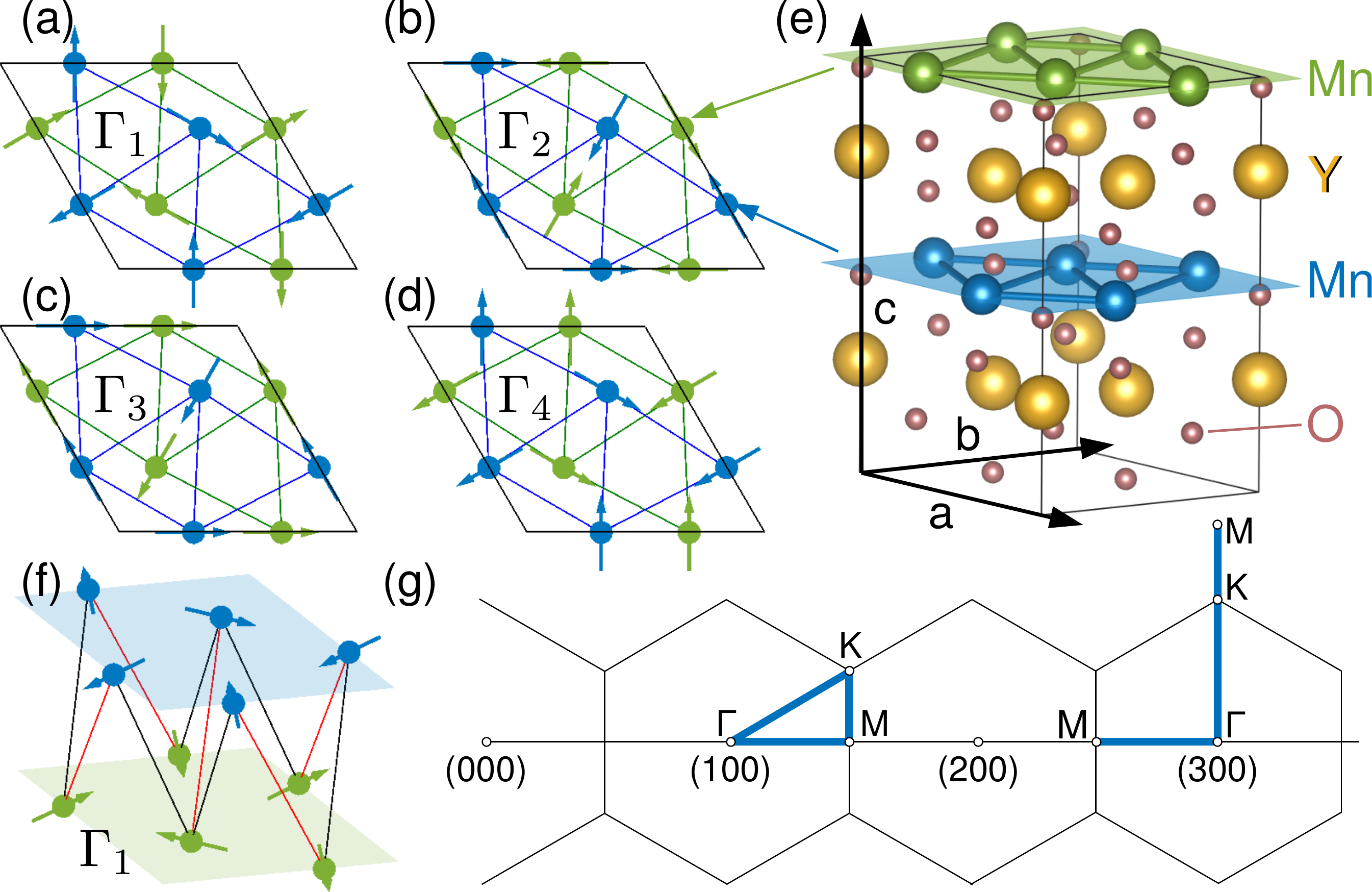}
\caption{
Panels (a-d) show four proposed antiferromagnetic ground states of h-YMnO$_3$ as seen along the $c$ axis. Their corresponding magnetic symmetry groups are P$6_3$cm ($\Gamma_1$), P$6_3$c'm' ($\Gamma_2$), P$6'_3$cm' ($\Gamma_3$), and P$6'_3$c'm ($\Gamma_4$). The lines indicate the nearest neighbor interaction, $J$. Blue and green colors indicate different Mn-O layers. (e) the Mn atoms in the simple unit cell are colored accordingly, as drawn by Vesta \cite{Momma2011}. (f) the inter-layer couplings are indicated with red, $J_{z_2}$, and black, $J_{z_1}$. (g) two different paths in reciprocal space in the $h$-$k$-plane used in our scattering experiments.
The blue lines indicate the main $q$-directions used for the scattering maps presented in Fig.~\ref{fig:maps}.
} \label{fig1_3D}
\end{figure}

An important class of multiferroics is the hexagonal rare-earth manganites $R$MnO$_3$, which are of type-I for $R$ being Sc, Y, Ho Er, Tm, Yb and Lu \cite{Petit2007,Sim2016}.
Due to its simplicity, with only one magnetic species, hexagonal (or h-) YMnO$_3$ is the most studied rare-earth manganite. 
The elementary cell of h-YMnO$_3$ is displayed in Fig.~\ref{fig1_3D}(e). 
The low magnetic ordering temperature of the material, $T_{\rm N}=72$ K, together with a high Curie-Weiss temperature, $T_{\rm CW}=-500$ K, provides a large frustration ratio of $f \approx 6.9$ \cite{Roessli2005,Lee2008}.

A giant magneto-elastic structural change has been reported in h-YMnO$_3$: Below $T_{\rm N}$, the Mn ions move from their symmetric positions, tripling the unit cell as a result \cite{Lee2008}. This observation was backed up theoretically \cite{varignon2013}, but has later been debated. The counter-agument is that due to overlapping magnetic and structural diffraction signals, the underlying Rietveld refinement could suffer from systematic errors\cite{Chatterji2012,Howard2013,thomson14}.

 In h-YMnO$_3$, the $S=2$ spins on the Mn$^{3+}$ ions order antiferromagnetically on triangles in the $a$-$b$-plane with a 120$^{\circ}$ angle between the neighboring spins \cite{Brown2006,Chatterji2007}. The 3D nature of the spin structure has recently been under intense debate. Using symmetry analysis, it was found that only the P6'$_{3}$cm' magnetic group would fit all observations \cite{Howard2013}. In contrast, a different group concluded that the magnetic order belongs to the magnetic P6'$_3$ space group, due to the observation of a small ferromagnetic component \cite{singh2013}.
 Four of the often investigated antiferromagnetic ground states are shown in Fig.~\ref{fig1_3D}.

\begin{figure*}[t]
\includegraphics[width=18cm]{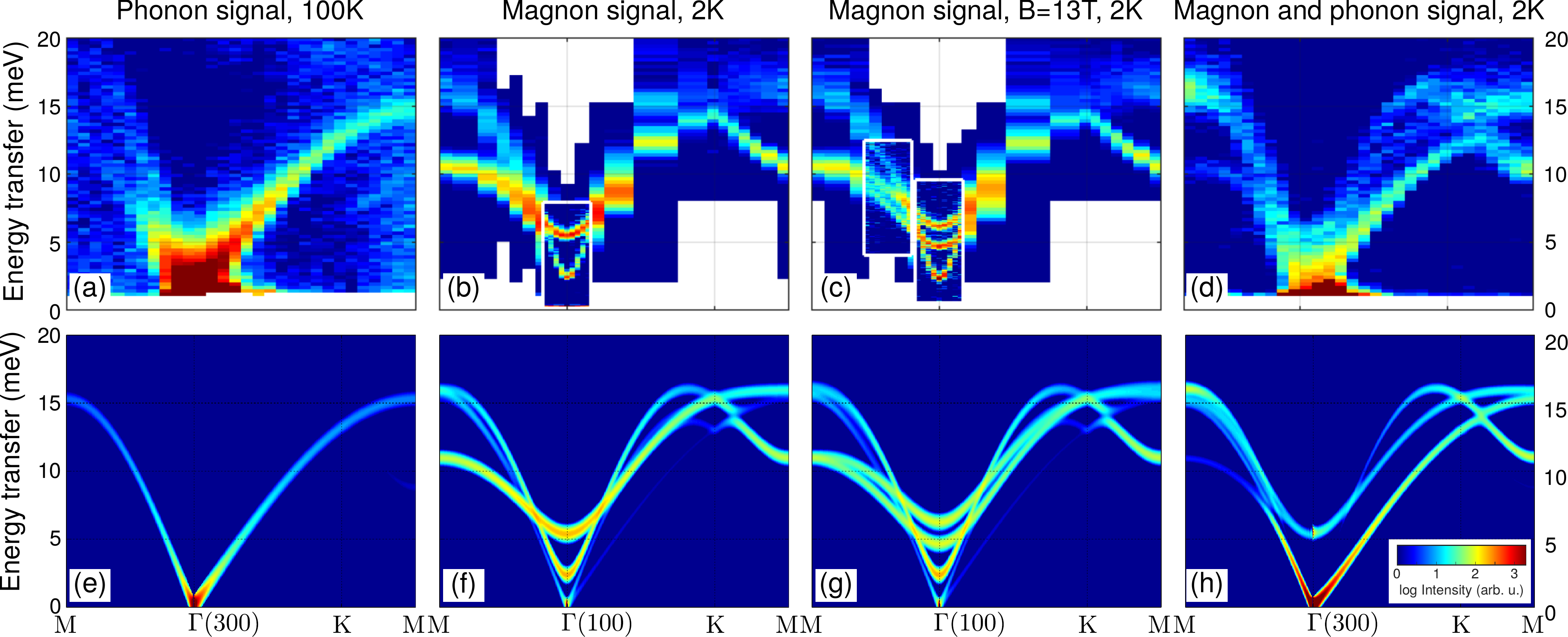}
\caption{
Color-maps of the magnon and phonon dispersions along the main symmetry directions in the $(a,b)$-plane in reciprocal space as indicated in Fig.~\ref{fig1_3D}(g). Panels (a-d) show our experimental results. (a) and (d) are measured in BZ(300) at 100~K and 2~K, respectively. (b) and (c) are measured in BZ(100) at 2~K and with an applied magnetic field of 0~T and 13~T, respectively. The high resolution data in (b) and (c) close to $\Gamma$ are measured at RITA-II, while the rest of the data are from EIGER; both instruments are located at PSI. Panels (e-h) show the theoretically calculated intensities of the dispersions at the same $q$-values, temperatures and applied magnetic fields using the most probable magnetic ground state. 
\label{fig:maps}}
\end{figure*}

The magnon modes \cite{Sato2003} and the magnon-phonon interaction \cite{Petit2007} in h-YMnO$_3$ were earlier studied with INS. 
The coupling between the two types of excitations was characterized with neutron polarization analysis and it was found that a hybridization occurs between the acoustic phonon and a magnon mode close to the zone center; at a value of the scattering vector of $\textbf{q}=(h\ 0\ 6)$ for $0<h<0.2$ \cite{Pailhes09}. 
Recently, the magnon dispersion in the full zone was measured in a single crystal and a powder phonon spectrum was modeled \cite{Oh2016}. From knowledge of the excitation spectra, assuming a classical $120^\circ$ 2D magnetic structure, a magnon-phonon interaction model was proposed. The non-linear terms in the Hamiltonian were found to cause a decay of the magnetic excitations. This model captures the measured magnon dispersion, but describes the obtained magnon intensities with limited accuracy. So far, no single model has been able to simultaneously describe the complexity of both magnons, phonons, and their interactions in h-YMnO$_3$.

In this work, we present single crystal INS measurements of magnons and phonons. We observe an avoided crossing at the zone boundary in the $(a,b)$ plane.
Neutron polarization analysis shows that the modes are of mixed magneto-structural character at this point. Furthermore, INS measurements with magnetic field along the $c$ direction reveal a linear splitting of the magnon in the entire zone, providing independent information about the 3D arrangement of the magnetic moments. Our theoretical model captures all experimental findings, including INS intensities. The model allows us to quantify the spin-lattice coupling and to identify P6$_3$cm and $P6_3c'm'$ as the two possible symmetries of the magnetic ground state.

Both the experimental results and theoretical modeling are presented in the main text below, while more details on the modeling can be found in the Appendix.

\section{Experimental Details}

%\label{Ap_A}
The sample used in the experiments were grown using the floating zone method \cite{Lancaster2007}. The crystal structure is hexagonal, with lattice constants $a=b=6.11$~\AA ; $c=11.39$~\AA . X-ray and neutron Laue investigations and neutron diffraction proved them mostly good single crystals with only a single phase and limited mosaicity. The crystals contained few completely misaligned grains, too small to contribute significantly to the inelastic scattering signal. Neutron diffraction was consistent with the lattice parameters and the magnetic ordering temperature (72 K), earlier reported for h-YMnO$_3$ \cite{Roessli2005}. The sample configuration and mount have been changed for the different experiments carried out to obtain the data presented in the paper. The experiments without an applied magnetic field were carried out on a single rod with a mass of 5.25~g. In order to fit the sample in the tight space of the cryo-magnet the sample had to be cut into two pieces. These were then co-aligned on top of each other. The data shown in Fig. \ref{inplane} were measured using a different piece (0.20~g) of sample with a larger mosaicity of 1.5$^{\circ}$

Inelastic scattering with cold neutrons was performed at the Paul Scherrer Institute (PSI), CH, using the \mbox{RITA-2} triple-axis spectrometer in the monochromatic imaging mode \cite{bahlrita1,bahlrita2}. All experiments were performed with a constant final energy of $5.0$~meV, with a Be-filter placed on the outgoing side. This gave an energy resolution of 0.2-0.5~meV, depending on the value of energy transfer. The experiment was performed with 80' incoming collimation and a natural outgoing collimation of 40' from the imaging mode.

Inelastic scattering with thermal neutrons without polarization analysis was performed on the triple axis instrument EIGER \cite{Stuhr2017} at PSI  with a constant final energy of 14.7~meV. Double focusing of the monochromator and horizontal focusing of the analyzer was used. A 36~mm thick pyrolytic graphite filter was placed between the sample and the analyzer in order to suppress higher order neutrons. 

At both spectrometers at PSI, we used either a liquid He cryostat or an Oxford 15 T split-coil vertical field cryo-magnet. The latter was used without its lambda stage, meaning that the maximum achievable field was 13 T.

INS with neutron polarization analysis was performed at the thermal triple axis spectrometer C5 at Chalk River Laboratories, Canada. During the experiment, the neutron polarization was directed along the scattering vector, ${\bf q}$. The non-spin-flip data therefore represent the phonon signal only, while the spin-flip signal is purely magnetic - given that the polarisation of the beam is perfect. For this particular experiment the measured flipping ratio was 13.8; a number that has been taken into account in our modeling. The experiment was performed using a constant final energy of 14~meV giving an energy resolution of 1-2 meV, depending on energy transfer.

\section{Experimental Results}
An overview of our main results is displayed in Fig.~\ref{fig:maps}, showing the magnon and phonon dispersions  along the high symmetry directions in reciprocal space, as indicated in Fig.~\ref{fig1_3D}(g).

The unperturbed acoustic phonon dispersion is measured above $T_N$, at $T=100$~K, in the $(3\ 0\ 0)$ Brillouin zone (or ``BZ(300)''),
see Fig.~\ref{fig:maps}(a). Scanning from $\Gamma$ over K to M, a clear signal from the transverse phonon is present. The diffuse intensity below the phonon branch between K and M is only observed above $T_N$ and is attributed to magnetic critical scattering. 
Fig.~\ref{fig:maps}(e) shows our model calculation of the neutron scattering intensity, as detailed in the theory section. 

Fig.~\ref{fig:maps}(b) displays the data obtained in BZ(100) at $T=2$ K. Due to the small nuclear dynamical structure factor at these low $\textbf{q}$-values the phonon cross section is negligible and the data shows a pure magnon signal. The corresponding result of our model is shown in Fig.~\ref{fig:maps}(f). 

\begin{figure}[tb]
\includegraphics[width=\columnwidth]{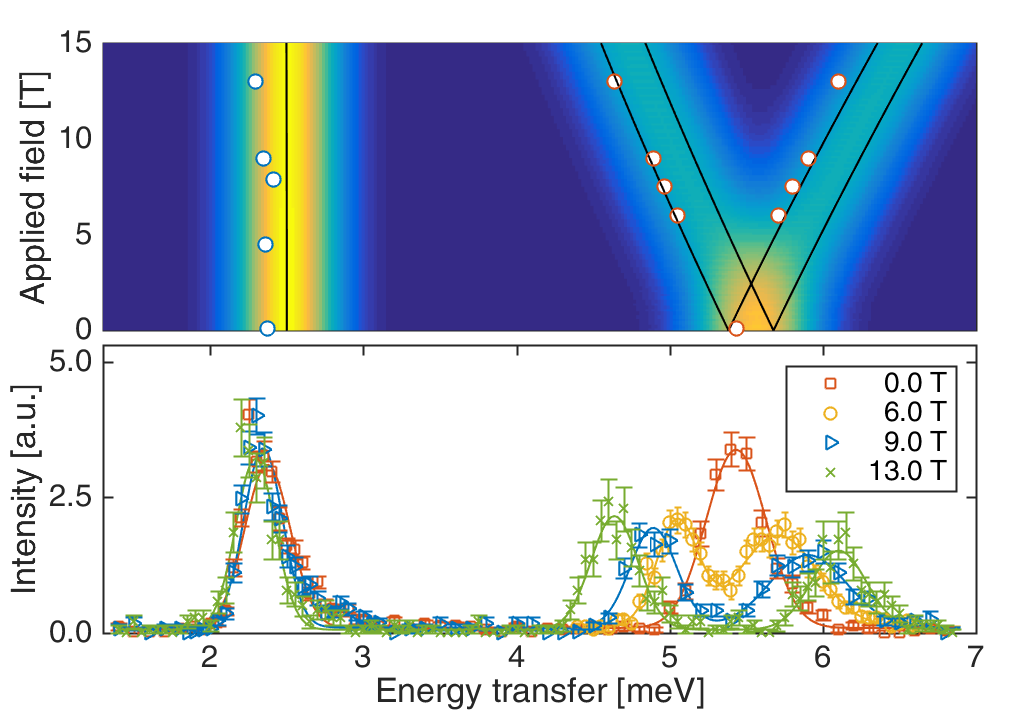}
\caption{The magnon energies at $\textbf{q}=(100)$ as a function of magnetic field along the $c$-direction. The colors represents the calculated intensities with the black lines being the magnon positions. The circles show the center positions of the experimental data, measured at 2 K. Bottom panel shows raw data with Gaussian fits. The data were measured at RITA-II, PSI. 
\label{fig:split}}
\end{figure}

With an applied magnetic field of 13 T along the $c$-axis, the degenerate upper magnon dispersions split. This is particularly clear close to $\Gamma$, as shown in Fig.~\ref{fig:maps}(c). Our model also captures this splitting, as shown in Fig.~\ref{fig:maps}(g). The complete field dependence of the splitting at the zone center in BZ(100) can be seen in Fig.~\ref{fig:split}. Due to the instrument resolution, the two peaks are only clearly distinguished at fields above 4.5~T. Both branches are seen to have a Zeeman-like linear field dependence. 
Our model captures this behavior, as seen in Fig.~\ref{fig:maps}(g). The split mode is really two almost-degenerate doublets that each split linearly with field. The difference between the modes is that one has the spin fluctuations in different planes parallel - the other has antiparallel fluctuations, leading to the difference in the energy shift in field. For comparison, the 2~meV mode is non-degenerate and is not affected by the field.

\begin{figure}[tb]
 \includegraphics[width=\linewidth]{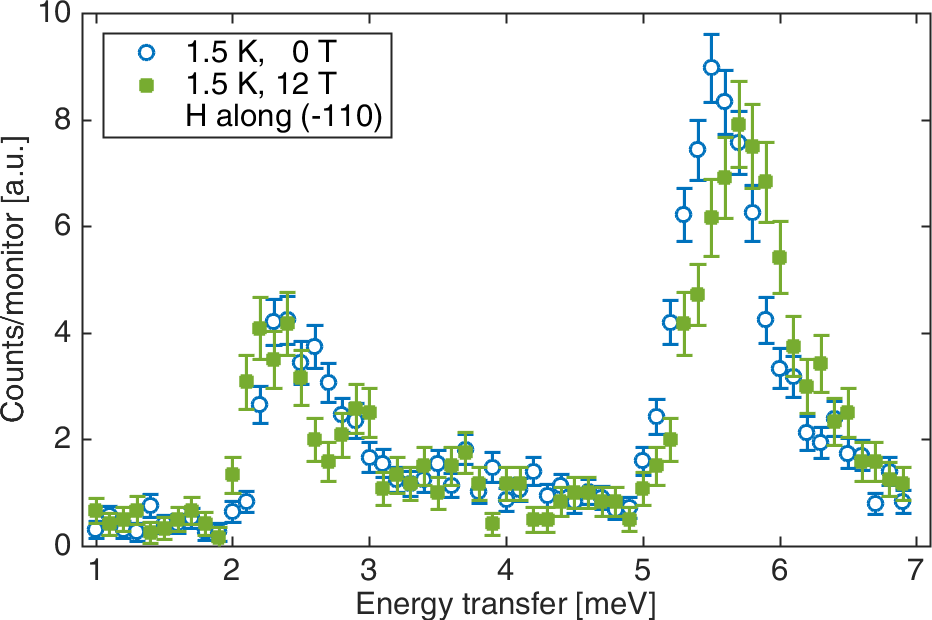}
\caption{ Raw neutron data of constant $q$ scan at ${\bf q}=(100)$ in zero magnetic field, and with an applied magnetic field of 12~T along the $(\bar{1}10)$ direction. The data were measured at EIGER, PSI.
\label{inplane}}
\end{figure}

A subtle effect of an applied magnetic field in the $a$-$b$-plane is observed as shown in Fig. \ref{inplane}. There are signs of a possible field-induced splitting of the lower magnon mode between 2 and 3~meV, while the upper magnon mode (that was found to split in a field along $c$) here only shifts to slightly higher energies. 
However, these measurements were performed with significantly worse instrumental resolution than those with field along the $c$-axis. Hence, more precise measurements are needed to draw firm conclusions on the effect of a field along this direction.

Fig.~\ref{fig:maps}(d) shows the magnon and phonon dispersions  at $T=2$~K in BZ(300). Here, the two signals are comparable in intensity. There is a clear avoided crossing at the K point, which indicates a pronounced coupling between the two types of excitations. This effect is also captured by our model as seen from Fig.~\ref{fig:maps}(h). 

\begin{figure}[tb]
\includegraphics[width=\columnwidth]{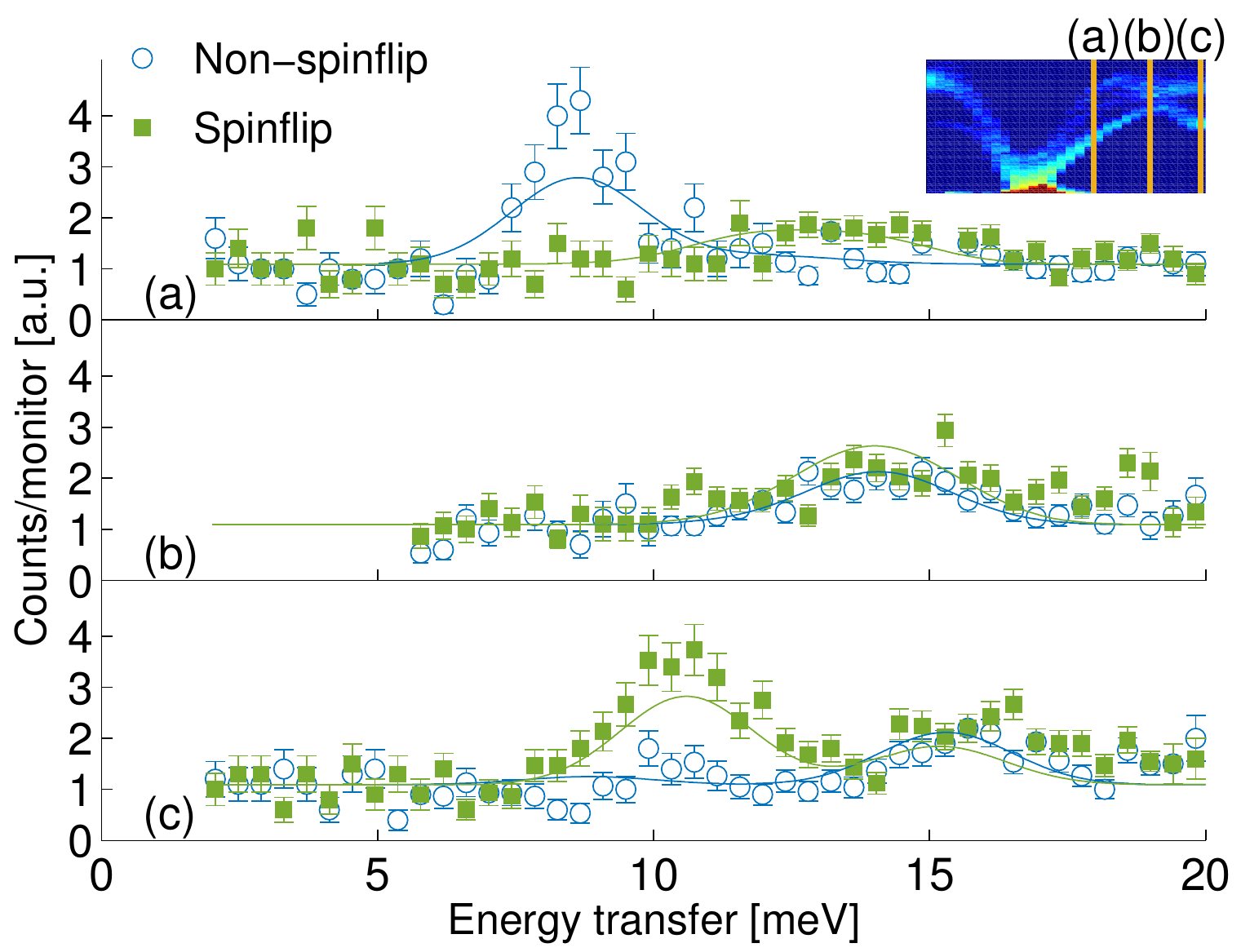}
\caption{Constant-${\bf q}$ scans at 40 K, measured with neutron polarization analysis. (a) $\textbf{q}=(2.8~0.4~0)$, (b) $\textbf{q}=(2.625~0.75~0)$, (c) $\textbf{q}_{\rm M}=(2.5~1~0)$. Blue and green lines represent calculations for the spin-flip and non spin-flip channels using Eq.~(\ref{eq_nuc_structure}) and Eq.~(\ref{eq_mag_structure}), respectively. In the inserted color map, the vertical orange lines indicate the positions of the three scans. The data were measured at C5, Chalk River Laboratories.
\label{fig:raw}
}
\end{figure}

To investigate the nature of the excitations, polarization analysis was performed in three constant-${\bf q}$ scans  close to the K point. The data are shown in Fig.~\ref{fig:raw} along with simulations of relative intensities. The non-spin-flip data capture the phonon signal, while the spin-flip signal is purely magnetic \cite{polarizationtextbook}. On the left side of the crossing, at $\textbf{q}=(2.8\ 0.4\ 0)$, a pure transverse acoustic phonon branch and a pure magnon branch can be distinguished. The data at the crossing, $\textbf{q}_{\rm K}=(2.67~0.67~0)$, do show a single branch with simultaneous signal in both spin-flip and non-spin-flip scattering. This could indicate either two modes of mixed magnon-phonon nature (merged due to limited energy resolution), or a mode crossing. However, we know from the data in Fig.~\ref{fig:maps}(d) that the latter possibility can be ruled out. Finally, at $\textbf{q}_{\rm M}=(2.5~1~0)$ , we see that the lower mode is of pure magnetic character, while the upper mode seems to be mixed magneto-structural.

\section{Theory}
We now outline the theoretical framework to describe coupled magnetoelastic excitations and describe the steps
needed to
model the INS data \cite{Lovesey_book}. 

\subsection{Spin Hamiltonian}
The starting point is a Hamiltonian of localized spins ${\bf S}_{i \bf R}$ with total spin $S=2$ at the Mn positions of h-YMnO$_3$.
Here, $\bf R$ labels the elementary cell and $i=1\ldots 6$ labels the positions within the cell, see Fig.\ \ref{fig1_3D}(e).
The spins interact by a Heisenberg interaction $J_{ij,\bf R  \bf R'}$, an easy plane-anisotropy $D$, and are subject to an effective site-dependent magnetic field ${\bf h}_i$
\begin{equation}
 H_S=\sum_{ij, \bf R \bf R'}J_{ij,{\bf R}  {\bf R}'} {\bf S}_{i\bf R}\cdot {\bf S}_{j {\bf R}'}+\sum_{i, \bf R}\bigl( {\bf h}_{i}\cdot {\bf S}_{i \bf R}+D S_{i\bf R}^z S_{i\bf R}^z\bigr)\;.\label{eq_HS}
\end{equation}
We consider a nearest neighbor Heisenberg coupling, $J$, in the plane and out-of-plane couplings, $J_{z_1}$ and $J_{z_2}$, which are inequivalent due to the dislocation of the Mn atoms from the $x=1/3$ positions, see Fig.~\ref{fig1_3D}(f).
The effective magnetic field is given by ${\bf h}_{i}={\bf h}-H{\bf m}_i$ where $\bf h$ is the external field, $H$ is an easy-axis anisotropy
and ${\bf m}_i$ is the direction of the local magnetization.
The spin operators $ {\bf S}_{i\bf R}$ are mapped to bosonic operators via a Holstein-Primakoff transformation. For this purpose, we find the classical ground state of the system and parametrize the local magnetization via a rotation angle. This is used to define a local coordinate system at each lattice point. Details on the calculations, which are standard within a spin-wave approach\cite{Toth15,Kreisel2008}, can be found in Appendix \ref{Ap_B}.

\subsection{Lattice Hamiltonian}
The lattice degrees of freedom are modeled in the harmonic approximation by a 
general phonon Hamiltonian of $s=1\ldots M$ modes written in terms of bosonic operators $a_{{\bf k},s}$ with eigenenergies $w_{{\bf k},s}$,
\begin{equation}
 H_L=\sum_{{\bf k},s} w_{{\bf k},s}a_{{\bf k},s}^\dagger a_{{\bf k},s}\,.
\end{equation}
To be specific, we restrict ourselves to the acoustic phonons in a hexagonal lattice which are describing the observed phonon modes in the absence of magnetic order in h-YMnO$_3$, see Appendix \ref{Ap_C}.

\subsection{Magnon-Phonon interaction}\label{Ap_D}
The coupling of magnons and phonons via the crystalline field can be described by
the Hamiltonian\cite{Laurence73,Boiteux72}
\begin{equation}
 H_{\text{SL}}=\sum_i \sum_{\alpha\beta\gamma\delta} G_{\alpha\beta\gamma\delta} S^\alpha_iS^\beta_i \epsilon_{\gamma\delta}^i
\end{equation}
where $G$ is the spin-phonon coupling tensor. We rewrite these in terms of irreducible representations $\sigma$ of the spin and lattice functions in the hexagonal symmetry class as\cite{Callen65}
\begin{equation}
 H_{\text{SL}}=-\sum_{i}\sum_{\sigma} B^\sigma(i) \sum_{i'}\epsilon_\sigma^{i'} S_{\sigma}^{i'}(i)\,,
\end{equation}
where the symmetry allowed couplings are $\vec B=[B^\alpha_{12},B^\alpha_{22},B^\gamma,B^\epsilon]$. The irreducible representations of the
strain tensor $\epsilon_{\gamma\delta}^i$ are linear combinations of the Cartesian strain tensor
\begin{equation}
 \epsilon_{\alpha\beta}^i=\frac 12 (E_{\alpha\beta}+E_{\beta\alpha})=\frac 12 \biggl(\frac{\partial X_i^\beta}{\partial r_\alpha}+\frac{\partial X_i^\alpha}{\partial r_\beta}\biggr),
 \label{eq_strain}
\end{equation}
and the spin tensors $S_{\sigma}^{i'}(i)$ are products of two components of the spin operators such that we can rearrange into
\begin{equation}
 H_{SL}=-\sum_{i\bf R} {\bf S}_{i\bf R}^T  E_i {\bf S}_{i\bf R}
\end{equation}
with the matrix\cite{Callen65}
\begin{widetext}
\begin{equation}
 E_i=\left(
 \begin{array}{ccc}
  B_{12}^\alpha \epsilon^{\alpha,1}-\frac{B_{22}^\alpha}{2\sqrt{3}}\epsilon^{\alpha,2}&\frac{B^\gamma}{2}\epsilon^\gamma_2&\frac{B^\epsilon}{2}\epsilon_2^\epsilon\\
  \frac{B^\gamma}{2}\epsilon^\gamma_2& B_{12}^\alpha \epsilon^{\alpha,1}-\frac{B_{22}^\alpha}{2\sqrt{3}}\epsilon^{\alpha,2}-\frac{B^\gamma}{2}\epsilon^\gamma_1&\frac{B^\epsilon}{2}\epsilon_1^\epsilon\\
  \frac{B^\epsilon}{2}\epsilon_2^\epsilon&\frac{B^\epsilon}{2}\epsilon_1^\epsilon& B_{12}^\alpha \epsilon^{\alpha,1}+\frac{B_{22}^\alpha}{\sqrt{3}}\epsilon^{\alpha,2}+\frac{B^\gamma}{2}\epsilon^\gamma_1
 \end{array}
 \right)
\end{equation}
\end{widetext}
and use Eq. (\ref{eq_hp_rotation}). 
The nonlocal contributions of the strain tensor can be obtained by a method introduced in Ref. \onlinecite{Evenson69}
where the local strain is replaced by a nearest neighbor contraction
\begin{equation}
\epsilon_{\alpha\beta}^i \rightarrow \tilde\epsilon_{\alpha\beta}^i=\frac 1 n \sum_{\bf \delta} \epsilon_{\alpha\beta}(i,i+\bf \delta)
\end{equation}
where ${\bf \delta}$ is the sum over nearest neighbors and $n$ is a normalization constant that ensures that
$\tilde\epsilon_{\alpha\beta}^i$ reduces to $\epsilon_{\alpha\beta}^i$ in the long-wavelength limit\cite{Jensen_risoe}.
The components of the matrix $E_i$ are given by
\begin{equation}
 E_i^{\alpha\beta}=\frac{1}{N}\sum_{{{\bf k}}, s} \frac i2 g_{{{\bf k}},s}^{\alpha \beta}\frac{a_{{{\bf k}},s}+a_{{{\bf k}},s}^\dagger}{\sqrt{2mw_{{{\bf k}},s}}} e^{i{{\bf k}}\cdot {{\bf R}}_i}\,,
\end{equation}
and the coupling constants $g_{{{\bf k}},s}^{\alpha \beta}$ are sums of products of $B^\sigma$ from the spin-lattice Hamiltonian and momentum-dependent structure 
function and the phonon polarization ${\bf g}\cdot {\bf e}_{{{\bf k}},s}$.
For the triangular lattice, we obtain the following structure function ${{\bf g}}({{\bf k}})=(g^{x},g^{y},g^{z})$:
\begin{subequations}
 \begin{align}
 g^{x}&=\frac 1{2a}\sin\biggl(\frac h 2\biggr)\cos\biggl(\frac h 6+\frac k 3\biggr)\\
 g^{y}&=\frac 1{2\sqrt{3}a}\biggl[\sin\biggl(\frac{h}{6}+\frac{k}{3}\biggr)\cos\biggl(\frac{h}{2}\biggr)+\sin\biggl(\frac{h}{3}+\frac{2k}{3}\biggr)\biggr]\\
     g^{z}&=\frac 1 c \sin l\,
  \end{align}
\end{subequations}
where the momentum $(hkl)$ is already expressed with respect to the relevant magnetic elementary cell.
Writing in momentum space,
\begin{equation}
 E_{i,{{\bf k}}}^{\alpha\beta}=\frac{4}{iS\sqrt S}\sum_s(a_{{{\bf k}},s}+a_{{{\bf k}},s}^\dagger)G_{{{\bf k}},s,i},
\end{equation}
we can see that the spin-lattice Hamiltonian is a hybridization term between the Holstein-Primakoff magnon operators and the phonon
operators with a product $G_{{{\bf k}},s,i}$ and the vectors of the rotated coordinate systems as matrix elements.

\subsection{Magnetoelastic waves}\label{Ap_E}

In summary, or model is given by
\begin{equation}
 H=H_S+H_{SL}+H_L
\end{equation}
which can be rearranged within linear spin-wave theory into the compact form
\begin{equation}
H= \sum_{\bf k}(\vec b_{\bf k}^\dagger,\vec b_{-\bf k})\mathcal D_{\bf k}\left(\begin{array}{c}
                                                                               \vec b_{\bf k}\\
                                                                               \vec b_{\bf -k}^\dagger
                                                                              \end{array}\right) \label{eq_Htot}\,.
\end{equation}
For the calculation of the ground state and the Fourier transforms of the terms from the spin Hamiltonian, we use the SpinW\cite{Toth15} package, and therefore
follow the notation for the matrices $A({{\bf k}})$, $B({{\bf k}})$ and ${\mathcal C}$ of that reference.
The grand dynamical matrix is then given by
\begin{equation}
 \mathcal D_{\bf k}=\left(
 \begin{array}{cccc}
  A({{\bf k}})-{\mathcal C} & \Gamma({{\bf k}}) & B({{\bf k}}) &\Gamma({{\bf k}})\\
  \Gamma^\dagger({{\bf k}}) & W({{\bf k}}) & \Omega({{\bf k}})&0\\
  B^\dagger({{\bf k}})& \Omega^\dagger({{\bf k}}) &\bar A(-{\bf k})-{\mathcal C} &\Omega({{\bf k}})\\
  \Gamma^\dagger({{\bf k}}) &0&\Omega^\dagger({{\bf k}}) &W(-{\bf k})
 \end{array}\right)\,,
\end{equation}
where the phonon dispersions appear in $W({{\bf k}})=\text{diag}(\{w_{{\bf k},s}\})$ and the elements of the matrices describing the magnon-phonon vertices are given by
\begin{subequations}
 \begin{align}
 \Gamma({{\bf k}})^{is}&={\bf e}_i^{-T}G_{{\bf k},s,i}{\bf m}_i\\
 \Omega({{\bf k}})^{is}&={\bf m}_i^TG_{-{\bf k},s,i}{\bf e}_i^{-}
 \end{align}
\end{subequations}
Following Colpa \cite{Colpa78}, we use the algorithm to diagonalize
the Bosonic Hamiltonian giving the para-unitary matrix ${\mathcal J}_{{\bf k}}^{-1}$
to diagonalize the Hamiltonian Eq. (\ref{eq_Htot})\cite{Serga12},
\begin{equation}
        {\mathcal J}_{{\bf k}}^{-1}\vec{\gamma}_{{\bf k}}^\dagger=  \vec b_{{\bf k}}^\dagger\,
        \label{eq_genbogoliubov}
\end{equation}
e.g.,
\begin{equation}
 {H}_2=\frac 12 \sum_{{\bf k}}\vec \gamma_{{\bf k}}^\dagger{\mathcal E}_{{\bf k}} \vec \gamma_{{\bf k}}+E_0^{(2)}
\end{equation}
with the diagonal matrix ${\mathcal E}_{{\bf k}}=\text{diag}(\{\omega_{{\bf k},l}\})$. In other words, ${\mathcal J}_{{\bf k}}$ is the wave function of the coupled magnetoelastic waves.
With our choice of the ordering in $\vec b_{\bf k}$, we can split up to the spin and lattice part of the wave function by
\begin{equation}
 {\mathcal J}_{{\bf k}}^{-1}=\left(
  \begin{array}{cccc}
\mathcal N^{\uparrow}\\
\mathcal M^{\uparrow}\\
\mathcal N^{\downarrow}\\
\mathcal M^{\downarrow}
  \end{array}\right)
\end{equation}
and define the matrices $\mathcal N$ and $\mathcal M$ via
\begin{equation}
\mathcal N=\left(
   \begin{array}{cccc}
    \mathcal N^{\uparrow}\\
\mathcal N^{\downarrow}
   \end{array}\right)\,,
   \quad
   \mathcal M=\left(
   \begin{array}{cccc}
    \mathcal M^{\uparrow}\\
\mathcal M^{\downarrow}
   \end{array}\right)\,.
   \label{eq_wave_function}
\end{equation}

\subsection{Dynamical structure factor}

The magnon part of the wave function ${\mathcal N}_{li}({\bf k})$ and the phonon part of the wave function ${\mathcal M}_{si}({\bf k})$ are given by the matrix elements of the matrices $\mathcal N$ and $\mathcal M$ as defined in Eq. (\ref{eq_wave_function}).
Using the general expression for the magnetic neutron scattering cross section, see Appendix \ref{Ap_magnetic_cross}, and inserting the magnon wave function, we obtain for the dynamical structure factor for magnetic INS
\cite{Toth15}
\begin{equation}
 { S}^{\alpha\beta}_{\text{mag}}({\bf q},\omega)=\sum_{l=1}^{2(N+M)}\left[\mathcal N^\dagger e^{\alpha\beta}({\bf k}) \mathcal N\right]_{ll} \Delta(\omega,{\bf q},l)\, .\label{eq_mag_structure}
\end{equation}
The matrix $e^{\alpha\beta}(\bf k)$ contains products of the components of the spherical unitary vectors defining the local coordinate system\cite{Toth15}, and 
{$\Delta(\omega,{\bf q},l)= \delta(\omega- \omega_{{\bf q},l})n(\omega)$
for $l\leq N+M$ and $\Delta(\omega,{\bf q},l)= \delta(\omega+ \omega_{{\bf q},l})[n(\omega) +1]$
for $l>N+M$ where $\omega_{{\bf q},l}<0$.}
As we derive in more detail in Appendix \ref{Ap_C} by inserting the phonon part of the wavefunction into the expression for the nuclear neutron scattering cross section\cite{Lovesey_book}, the dynamical structure factor for nuclear INS is given by
\begin{align}
 S_{\text{nuc}}({\bf q},\omega) =\sum_{s} |{\bf q}\cdot{\bf e}_s|^2
  \sum_{l=1}^{2(N+M)}\frac{\mathcal M_{si}}{mw_{{\bf q}s}}\Delta(\omega,{\bf q},l)\, ,\label{eq_nuc_structure}
\end{align}
where ${\bf e}_s$ is the polarization vector of the phonon mode $s$, and $m$ is the mass of the atoms.

It turns out that a minimal model of 6 magnetic ions in the elementary cell \cite{Petit2007,Toulouse2014} together with the 3 acoustic phonon modes
is sufficient to explain all our experimental findings up to 20~meV energy transfer.
The geometry, the considered ground states, and the in-plane couplings are shown in Fig.~\ref{fig1_3D}.
The layers can couple ferromagnetically or antiferromagnetically as revealed by a symmetry analysis \cite{Howard2013}. 
The Heisenberg couplings as well as the anisotropies are free parameters within our theory and have been fitted to yield agreement with our experimental data for each of the magnetic ground states shown in Fig.~\ref{fig1_3D}.
{We have also considered linear combinations of two pairs of ground state configurations\cite{Fabreges_PhD} which did not yield a better agreement between theory and experiment.}

To be more quantitative, a fit of the acoustic phonon bandwidth as one free parameter for the lattice vibrations gives good agreement with the measured spectra at 100~K, yielding $\sqrt{C/m}=6.05(5)\,\text{meV}$, see Appendix \ref{Ap_B}. At low temperatures, this value is $6.25(9)\,\text{meV}$, due to hardening of the crystal.
Next, we calculate the transverse part of the spin dynamical structure factor (including the magnetic form factor of a Mn$^{3+}$ ion \cite{TabCrys}),
since it is proportional to the measured neutron signal \cite{Lovesey_book}.
The optimized model parameters are $J=2.43(2)\,\text{meV}$, $D=0.32(2)\,\text{meV}$, $H=0.49(4)\,\mu \text{eV}$ and for the ferromagnetic out-of-plane couplings $J_{z_1}=-150.9(6)\,\mu \text{eV}$, and $J_{z_1}-J_{z_2}=-2.4(2)\,\mu \text{eV}$.
The values of the symmetry-allowed elastic coupling constants are $[B^\alpha_{22},B^\gamma,B^\epsilon]=[19(4),15(3),10(2)]\,\text{meV}^{3/2}a\sqrt m$, (see Appendix \ref{Ap_magnetic_cross}).

\section{Discussion}
Experimentally, we observe an avoided crossing of the magnon and phonon branches in the $(a,b)$-plane, at the boundary of the Brillouin zone (the $K$-point). This complements earlier reports on a similar crossing closer to the zone center along the $c$-direction \cite{Petit2007,Pailhes09}. Both findings underline the significance of the magneto-elastic coupling in h-YMnO$_3$. 
\begin{figure*}[tb]
 \includegraphics[width=\linewidth]{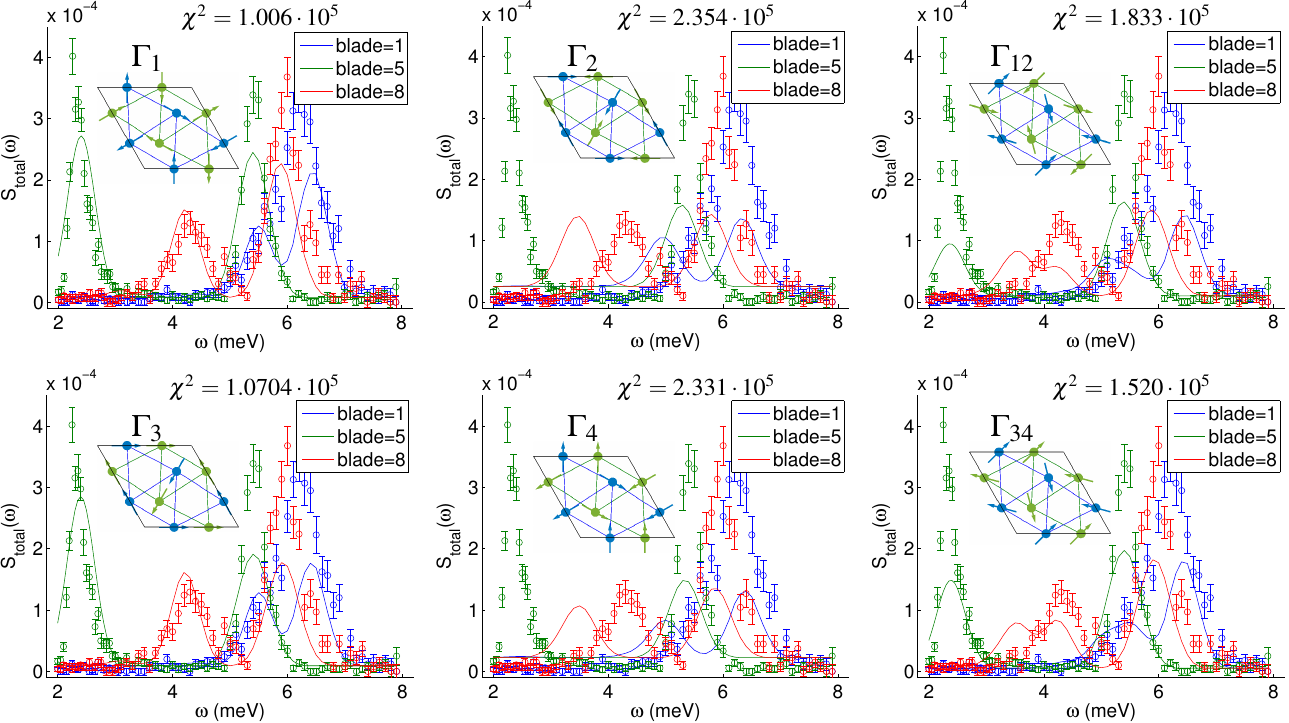}
\caption{Measured spectra (error bars) at $T=2\,\text{K}$ at selected momenta in direct comparison to the full model in the ordered state for fits to calculations assuming 6 different ground states. The corresponding spin structures are shown in the respective insets.
\label{fig_fit_mag}}
\end{figure*}
We discovered a clear Zeeman-like splitting of the 5~meV magnon mode with a magnetic field applied along the $c$-axis. The observed symmetric splitting cannot be explained by a pure 2D model or by a number of the possible 3D models of the magnetic ground state. Hence, the splitting has been used to obtain solid information on the 3D nature of the magnetic structure. The magnetic field dependence of the magnon mode was earlier studied by optical spectroscopy \cite{Toulouse2014}, revealing, curiously, only the upper branch of the split modes. We speculate that this could be due to the selection rules of the optical spectroscopy, combined with the difference in c-axis polarization of the magnon modes. 

Combining the Zeeman splitting with the information on the measured magnon and phonon intensities, we are able to exclude two of the magnetic ground states, $\Gamma_2$ and $\Gamma_4$ (see Fig.~\ref{fig1_3D}). The two other ground states, $\Gamma_1$ and $\Gamma_3$, are both overall compatible with the observations,  This is in agreement with the results from neutron powder diffraction \cite{Park2002}, where the two states are homometric and thus cannot be distinguished \cite{Howard2013}. By polarized neutron diffraction, it was concluded that $\Gamma_3$ (corresponding to P6'$_3$c'm) was close to the correct ground state, but most probably the spins were turned by approximately $11^\circ$ with respect to this state \cite{Brown2006}. Likewise, based upon the selection rules of second harmonic generation, it was earlier concluded that $\Gamma_3$ was the true ground state \cite{Fiebig2000}.
Our INS data give independent evidence that the true ground state is either P6$_3$cm or P6'$_3$c'm, and our modeling shows that these states are not homometric in the inelastic channel. However, our data is not of sufficient quality to uniquely select one of the two states.

The magnon-phonon interaction was recently modeled by Oh {\it et al.}~\cite{Oh2016}, although they did not observe the phonon dispersion directly. Their model was based upon a 2D magnetic ground state. We believe this to have caused the observed discrepancies between their model and the measured magnon intensities.
In contrast, the present model, using either of the two possible 3D ground states of the magnetic system, gives a much better account of the measured magnon and phonon spectra and quantitatively model the magnon-phonon coupling, with agreement in both dispersions and intensities. Hence, we believe that our model for the low energy structural and magnetic dynamics in h-YMnO$_3$ is essentially correct.

\section{Conclusion}
We have observed a strong magneto-elastic coupling  in h-YMnO$_3$, leading to mixed magneto-structural excitations at the zone boundary in the $(a,b)$-plane. In addition, we have observed a linear field-induced splitting of the magnon dispersion, which in turn has led us to give an independent suggestion for the 3D magnetic ground state of the system to be of either the $\Gamma_1$ or $\Gamma_3$ type. Using either of these as the ground state, we can model the magnon-phonon interaction and reproduce both the observed dispersion relations and intensities accurately in the full Brillouin zone. 
Our results underline the importance of using the correct 3D ground state for modeling the otherwise predominantly two-dimensional Mn spin system. For this reason, and because h-YMnO$_3$ is the most simple of the hexagonal manganites, our results are of general relevance for the understanding of magnetism and magneto-elastic coupling in multiferroic materials.

\section*{Acknowledgements}
We thank Pia J. Ray and students from Univ.\ Copenhagen for assistance with experiments, Jens Jensen, Xavier Fabreg\`es, and Shantanu Mukherjee for stimulating discussions, and Tom Fennell for valuable comments to the manuscript. The project was supported by the Danish Research Council for Nature and Universe through ``Spin Architecture'' and ``DANSCATT''.
T.K.S.\ thanks the Siemens Foundation for support. A.K.\ and B.M.A.\ acknowledge support from Lundbeckfond fellowship (grant A9318).
This work is based on experiments performed at SINQ,
Paul Scherrer Institute (CH) and at Chalk River Laboratories (CAN).

\appendix

\section{Spin-wave expansion}\label{Ap_B}
As discussed in the main text, we start from a localized spin Hamiltonian with spin $S=2$ operators
${\bf S}_{i\bf R}$ located at the positions of the Mn atoms in the crystal structure %P6$_3$cm group 
with setting $x=0.315$\cite{Petit2007,Toulouse2014},
\begin{equation}
 H_S=\sum_{ij, \bf R \bf R'}J_{ij,{\bf R}  {\bf R}'} {\bf S}_{i\bf R}\cdot {\bf S}_{j {\bf R}'}+\sum_{i, \bf R}\bigl( {\bf h}_{i}\cdot {\bf S}_{i \bf R}+D S_{i\bf R}^z S_{i\bf R}^z\bigr)\;,\label{eq_HS1}
\end{equation}
with in-plane coupling $J_{ij,{\bf R}  {\bf R}'}=J$ if $(i,{\bf R})$ and $(j,{\bf R}')$ are nearest neighbors, and two non-equal out-of plane couplings
$J_{z_1}$ and $J_{z_2}$.

The in-plane exchange interactions are not equal by symmetry as well due to the deviations from the perfect $x=1/3$ positions of the Mn atoms. Taking this into account in the modeling would not reveal any new information because a small difference in the non-equivalent in-plane exchange couplings does not result in a qualitatively different behavior of the spin-wave modes at any point in the Brillouin zone and the small quantitative difference cannot be detected within the experimental resolution. Hence, we have not modeled this potential small in-plane coupling difference.

The matter is different for the out-of plane couplings: While the exact values of these couplings as well are difficult to fix from a fitting procedure, the spectra show a qualitatively different behavior at some points of the Brillouin zone if they are nonzero. The sign of the out-of plane couplings then selects the ground state and leads to different nature of the eigenmodes which is visible in the scattering intensities.

The easy plane-anisotropy $D$ forces the spins in the classical ground state into the plane.
We write the easy-axis anisotropy $H$ in terms of an effective magnetic field, ${\bf h}_{i}={\bf h}-H{\bf m}_i$ where ${\bf{ h}}=h{\bf{ e}}_z$ is an external magnetic field and ${\bf m}_i$ is the direction of the local magnetization at site $i$. 
For calculation purposes, we use a non-symmetric effective g-tensor $g_i$ and we can express the effective magnetic field in terms
of an magnetic induction that is arbitrarily directed parallel to the crystallographic $c$ direction, ${\bf b}={\bf e}_z$, via ${\bf h}_i=g_i{\bf b}$.

Next, the classical ground state is determined by replacing the spin-operators in the above expression by ${\bf S}_{i\bf R}=S{\bf m}_i$ and parametrization of
the local coordinate system $\{ {\bf m}_i,{\bf e}_i^{(1)} ,   {\bf e}_i^{(2)}\}$\cite{Spremo2005,Kreisel2008}.
Introduction of the
spherical vectors ${\bf e}_i^{\pm}={\bf e}_i^{ (1)}\pm i{\bf e}_i^{(2)}$
allows us to rewrite this rotation as
 \begin{equation}
  {\bf S}_{i\bf R}=S_{i\bf R}^\parallel {\bf m}_i+\frac 12 \sum_{p=\pm}S_{i\bf R}^{-p} {\bf e}_i^{p}.
 \end{equation}
With the Holstein-Primakoff transformation (up to leading order in $1/S$)
\begin{subequations}
  \begin{align}
S_{i\bf R}^+\approx\sqrt{2S}b_{i\bf R}\,,\\
S_{i\bf R}^-\approx\sqrt{2S} b_{i\bf R}^\dagger\,,\\
S_{i\bf R}^\parallel=S-b_{i\bf R}^\dagger b_{i\bf R}\,,\label{eq_HP}
  \end{align}
  \end{subequations}
we obtain the form
\begin{equation}
 {\bf S}_{i\bf R}=\sqrt{\frac S2}\bigl({\bf e}_i^{-} b_{i\bf R}+{\bf e}_i^+b_{i\bf R}^\dagger\bigr)+{\bf m}_i (S-b_{i\bf R}^\dagger b_{i\bf R})\label{eq_hp_rotation}
\end{equation}
such that all coefficients for the magnon operators $b_{i\bf R}$ in the quadratic Hamiltonian can be collected straight forward\cite{Toth15} and transformed to momentum space.
\begin{figure}[tb]
  \includegraphics[width=\linewidth]{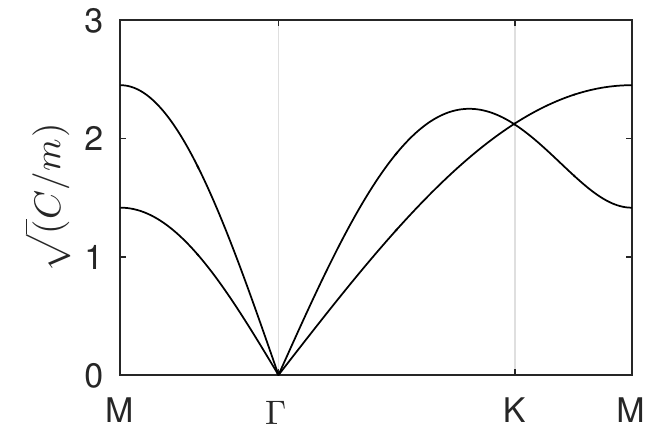}
 \caption{Phonon dispersion plotted along high symmetry directions\cite{Martinsson03}.}
 \label{fig_phonon_spectrum_hex}
 \end{figure}
\begin{figure}[tb]
 \includegraphics[width=\linewidth]{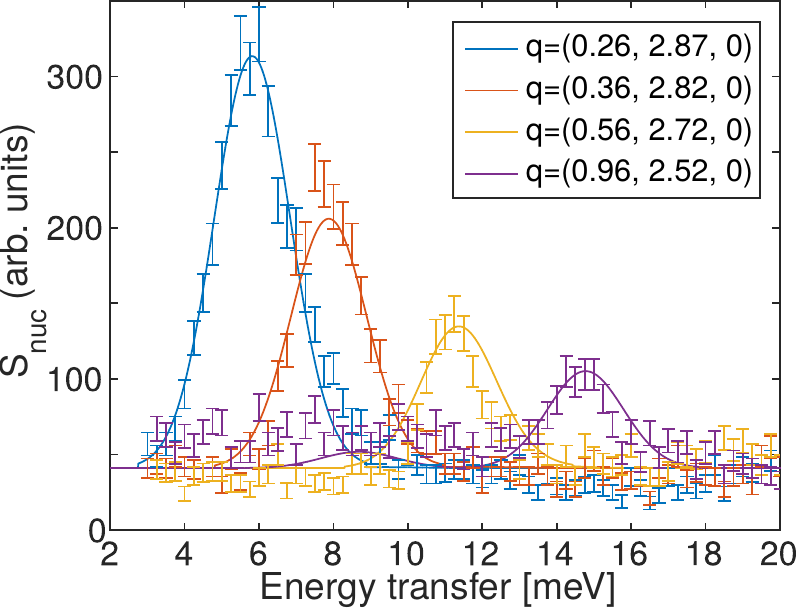}
\caption{Measured phonon spectra (error bars) at $T=100\,\text{K}$ at selected momenta in direct comparison to the model of acoustic phonons (solid lines) on a triangular lattice, Eq. (\ref{eq_HL}), calculated using Eq. (\ref{eq_nuc_structure1}) setting $\sqrt{C/m}=6.05\,\text{meV}$.
\label{fig_fit_phonon}}
\end{figure}

\section{Phonon modes in a triangular lattice}\label{Ap_C}

To calculate the phonon modes from a simple model, we use a triangular lattice of Mn atoms (for simplicity located at the
ideal $x=1/3$ positions) with lattice constant $a$, mass $m$ and the positions of the
corresponding Bravais lattice $\{ {\bf{R}}_i \}$ coupled to their nearest neighbors with a spring constant $C$.
Writing down the equations of motion we obtain the eigenfrequencies $w_{{\bf{k}},s}$ and the corresponding
eigenvectors ${\bf e}_{{{\bf k}},s}$ via a normal mode analysis and write the  
Hamiltonian describing lattice vibrations\cite{Martinsson03}, 
\begin{equation}
{H}_{L} =   \sum_{ {{\bf k}}  s } w_{{{\bf k}},s } \left( a^{\dagger}_{{{\bf k}}, s}
 a_{{{\bf k}}, s} + \frac{1}{2} \right),
 \label{eq_HL}
 \end{equation}
where $a_{\bf{k},s}$ annihilates a phonon with wave-vector $\bf{k}$ and
polarization $s$.
In our system, the phonon modes are modeled as three acoustic modes $s=1,2,3$, two of them obtained from the two
dimensional system as discussed above, the third obtained from a rotation of the polarization vector ${\bf e}_{{\bf k},s}$ of the transverse mode out of the plane by keeping
the eigenenergy degenerate. A plot of the phonon dispersions along high symmetry directions is shown in Fig. \ref{fig_phonon_spectrum_hex}.
We now have the dynamic positions of the atoms in the full elementary cell ${{\bf r}}_i = {{\bf R}}_i + {{\bf X}}_i$.
The lattice distortions in momentum space are quantized in the usual way,
 \begin{align}
 {{\bf X}}_{{{\bf k}}}& =  \sum_{s}   \frac{{\bf e}_{ {{\bf k}},s}}{ \sqrt{2 m w_{{{\bf k}} s }} }
 ( a_{ {{\bf k}} s } +  a_{ -{{\bf k}} s }^{\dagger} ) .
 \label{eq_phononop}
 \end{align}

\begin{figure}[tb]
 \includegraphics[width=\linewidth]{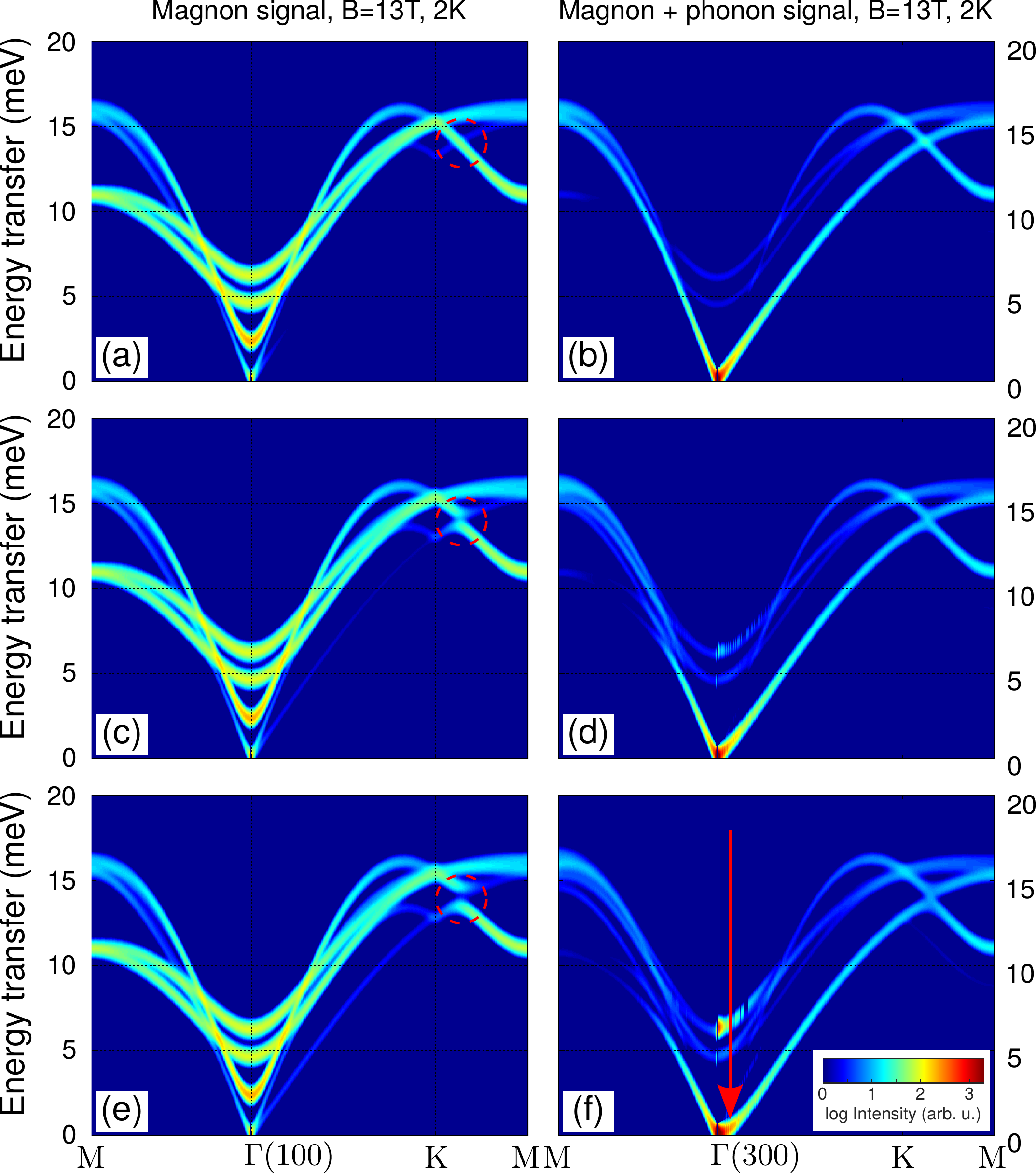}
\caption{Calculation of the total INS cross section along the experimental momentum path around $(1 0 0)$ and $(3 0 0)$ without magnon-phonon coupling (a,b), with the magnon-phonon coupling
as deduced in the main text (c,d) and a magnon-phonon coupling enhanced by 50\% (e,f). On increasing the magnon-phonon coupling, one can see the opening of a gap close to the K-point (red circle) and a softening of the phonon mode until it is very close to an instability (arrow).
\label{fig_compare_Bc}
}
\end{figure}

\begin{figure}[t]
 \includegraphics[width=\linewidth]{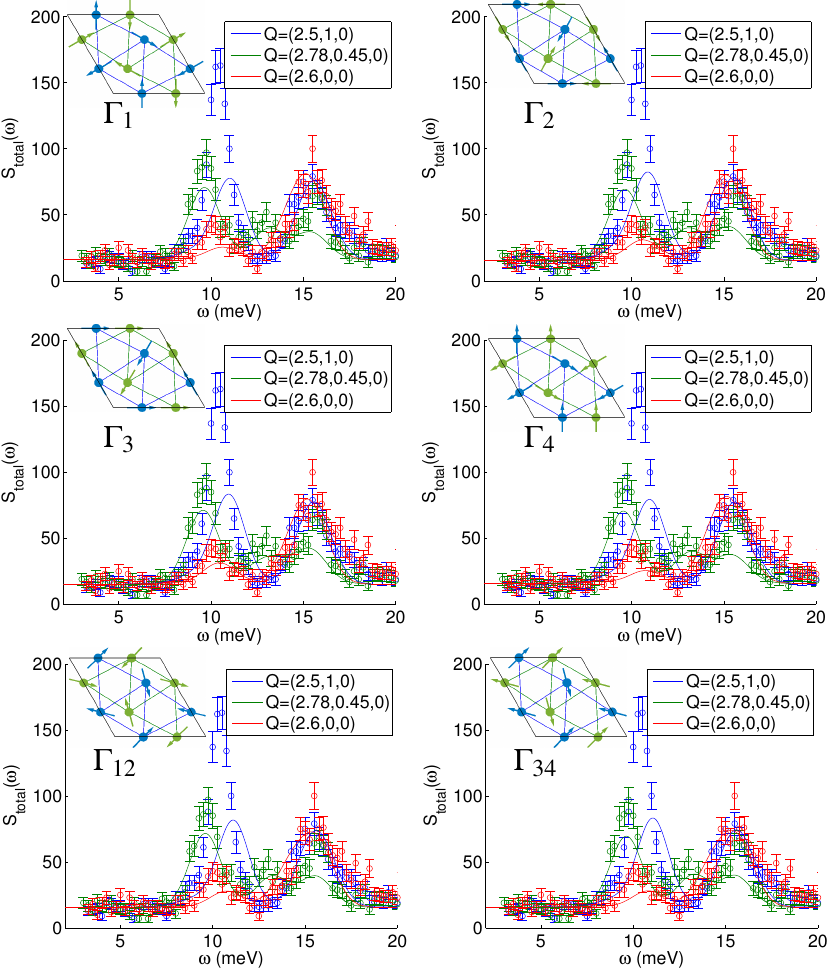}
\caption{Measured spectra (error bars) at $T=2\,\text{K}$ at selected momenta in direct comparison to the full model in the ordered state. The EIGER data represents scattering with defined momentum transfer ${\bf q}$ as indicated. The data shown cannot be used to distinguish the different considered ground states as the modeled intensities are almost identical.
\label{fig_fit_magnon_phonon}}
\end{figure}
\label{sec_mag}
\section{Nuclear neutron scattering cross section}\label{Ap_F}
In this section we derive the nuclear INS cross section in presence of magnetoelastic waves as derived above.
The starting point is the expression for the coherent nuclear INS cross section in Ref.~\onlinecite{Lovesey_book}, chapter 4.4:
\begin{align}
  \left.\frac{d^2 \sigma}{d \Omega dE'}\right|_{\text{coh}}^{\text{inel}}&=\frac{\sigma_i}{4\pi}\frac{k}{k'} \frac{1}{2\pi \hbar}\int_{-\infty}^{\infty} dt e^{-i\omega t} e^{-2W({{\bf k}})}\notag\\
 &\times \sum_{ij}e^{i{\bf k}\cdot({\bf R}_i-{\bf R}_j) \langle {\bf k}\cdot{\bf X}_i\,{\bf k}\cdot{{\bf X}}_j(t)\rangle}\,.
\end{align}
We know the time dependence of the magnetoelastic operators
\begin{equation}
 \gamma_l({\bf q},t)^\dagger=\gamma_l({\bf q})e^{i\omega_{{\bf q} l}t} 
\end{equation}
such that we use Eq. (\ref{eq_genbogoliubov}) to evaluate the expectation value $\langle {\bf k}\cdot{\bf X}_i\,{\bf k}\cdot{\bf X}_j(t)\rangle$.
Finally, we arrive at the expression
\begin{align}
  \left.\frac{d^2 \sigma}{d \Omega dE'}\right|_{\text{coh}}^{\text{inel}}=
  P_{\text{nuc}}\sum_{l,s}
  |{\bf k}\cdot{\bf e}_{{\bf k},s}|^2%\notag\\
  \frac{\mathcal M_{sl}}{mw_{{\bf k}s}} \Delta(\omega,{\bf k},l)\label{eq_nuc_structure1}
\end{align}
with
$\Delta(\omega,{\bf k},l)= \delta(\omega- \omega_{{\bf k},l}d_{l,l})[n(\omega) +\frac 12 (1-d_{ll})]$, $d_{ll}$ is a generalized Kronecker delta being negative for $l>N+M$ and $n(\omega)$ is the Bose function.
The prefactor $P_{\text{nuc}}= \frac{\sigma_c}{4\pi}\frac{k'}{k} \frac{(2\pi)^3}{2V_0M} e^{-2W({\bf q})}$ contains the Debye-Waller factor, the nuclear cross section of the corresponding atom and the kinematic factor $k'/k$. It is omitted in the main text,
since it is a momentum-independent constant which has to be adjusted to the experimental data.
When evaluating the expression above, we broaden the result in energy by a convolution with a Gaussian containing the experimental resolution.
\\

\section{Magnetic neutron scattering cross section}\label{Ap_magnetic_cross}

For the magnetic INS cross section, we follow the standard procedure in calculating the dynamical structure factor from the magnon
operators in the Holstein-Primakoff basis\cite{Lovesey_book}, given the eigenstates of the full Hamiltonian.
Performing all transformations, one arrives at the result\cite{Toth15}
\begin{equation}
 { S}^{\alpha\beta}_{\text{mag}}({\bf q},\omega)=\sum_{l=1}^{2(N+M)}\left[\mathcal N^\dagger e^{\alpha\beta}({\bf q}) \mathcal N\right]_{ll} \Delta(\omega,{\bf q},l)\, .
\end{equation}
For the total magnetic INS cross section
we use the expression
\begin{align}
  \left.\frac{d^2 \sigma}{d \Omega dE'}\right|_{\text{coh}}=P_{\text{mag}}f({\bf q})^2\sum_{\alpha\beta}(\delta_{\alpha\beta}-\frac{q_\alpha q_\beta}{{\bf q}^2}){ S}^{\alpha\beta}_{\text{mag}}({{\bf q}},\omega)\,,\label{eq_mag_structure1}
\end{align}
where we use the magnetic form factor $f({\bf q})$ of Mn$^{3+}$\ \cite{TabCrys}
and a experiment-specific prefactor $P_{\text{mag}}$ containing 
cross section prefactors and the kinematic prefactors.

We have considered four magnetic ground states as proposed earlier\cite{Howard2013}, as shown in Fig.~\ref{fig1_3D}. As discussed in the main text, two of the states are not compatible with the measured intensities, {\em e.g.} even a fit with allowing a change of all model parameters could not give a reasonable agreement, see Fig.~\ref{fig_fit_mag}.

As first step, we compare the model for the phonons to the measured spectra at $T=100$~K, e.g. above the magnetic ordering temperature.
In this case, the Bose-factor in 
Eq.~(\ref{eq_nuc_structure1}) enhances the low energy intensities significantly. While in the main text, we show the full data along the cut in the Brillouin zone as a color plot, we present
in Fig. (\ref{fig_fit_phonon}) a direct comparison of the spectra revealing that our simple model of acoustic phonons is sufficient to describe the lattice excitations at low energies.

In the magnetically ordered state, our model includes the parameters of two additional terms $H_S$ and $H_{SL}$. In order to fix them, we first minimize the difference between calculated
and measured spectra at momenta where $H_{SL}$ does not perturb the magnon modes, e.g. at the $\Gamma$ point to get an estimate of the magnon-only model parameters, then include further momenta where significant hybridization takes place to also
fix the magnon-phonon couplings $\vec B=[B^\alpha_{12},B^\alpha_{22},B^\gamma,B^\epsilon]$, see Fig. \ref{fig_fit_magnon_phonon}. Note that $B^\alpha_{12}$ does not enter the result and will not be considered further, and the values of the magnon-phonon couplings are large such that the system appears to be close to an instability as demonstrated in Fig. \ref{fig_compare_Bc}.
Note that in order to reproduce the structure of magnon excitations in the magnetic field, it is required to take into account all 6 magnetic ions in the elementary cell, e.g, also the interlayer coupling is needed.


\begin{thebibliography}{49}%
\makeatletter
\providecommand \@ifxundefined [1]{%
 \@ifx{#1\undefined}
}%
\providecommand \@ifnum [1]{%
 \ifnum #1\expandafter \@firstoftwo
 \else \expandafter \@secondoftwo
 \fi
}%
\providecommand \@ifx [1]{%
 \ifx #1\expandafter \@firstoftwo
 \else \expandafter \@secondoftwo
 \fi
}%
\providecommand \natexlab [1]{#1}%
\providecommand \enquote  [1]{``#1''}%
\providecommand \bibnamefont  [1]{#1}%
\providecommand \bibfnamefont [1]{#1}%
\providecommand \citenamefont [1]{#1}%
\providecommand \href@noop [0]{\@secondoftwo}%
\providecommand \href [0]{\begingroup \@sanitize@url \@href}%
\providecommand \@href[1]{\@@startlink{#1}\@@href}%
\providecommand \@@href[1]{\endgroup#1\@@endlink}%
\providecommand \@sanitize@url [0]{\catcode `\\12\catcode `\$12\catcode
  `\&12\catcode `\#12\catcode `\^12\catcode `\_12\catcode `\%12\relax}%
\providecommand \@@startlink[1]{}%
\providecommand \@@endlink[0]{}%
\providecommand \url  [0]{\begingroup\@sanitize@url \@url }%
\providecommand \@url [1]{\endgroup\@href {#1}{\urlprefix }}%
\providecommand \urlprefix  [0]{URL }%
\providecommand \Eprint [0]{\href }%
\providecommand \doibase [0]{http://dx.doi.org/}%
\providecommand \selectlanguage [0]{\@gobble}%
\providecommand \bibinfo  [0]{\@secondoftwo}%
\providecommand \bibfield  [0]{\@secondoftwo}%
\providecommand \translation [1]{[#1]}%
\providecommand \BibitemOpen [0]{}%
\providecommand \bibitemStop [0]{}%
\providecommand \bibitemNoStop [0]{.\EOS\space}%
\providecommand \EOS [0]{\spacefactor3000\relax}%
\providecommand \BibitemShut  [1]{\csname bibitem#1\endcsname}%
\let\auto@bib@innerbib\@empty
%</preamble>
\bibitem [{\citenamefont {Hill}(2000)}]{Hill2000}%
  \BibitemOpen
  \bibfield  {author} {\bibinfo {author} {\bibfnamefont {Nicola~A.}\
  \bibnamefont {Hill}},\ }\bibfield  {title} {\enquote {\bibinfo {title} {Why
  are there so few magnetic ferroelectrics?}}\ }\href {\doibase
  10.1021/jp000114x} {\bibfield  {journal} {\bibinfo  {journal} {J. Phys. Chem.
  B}\ }\textbf {\bibinfo {volume} {104}},\ \bibinfo {pages} {6694--6709}
  (\bibinfo {year} {2000})}\BibitemShut {NoStop}%
\bibitem [{\citenamefont {Lee}\ \emph {et~al.}(2008{\natexlab{a}})\citenamefont
  {Lee}, \citenamefont {Ratcliff}, \citenamefont {Cheong},\ and\ \citenamefont
  {Kiryukhin}}]{Lee2008a}%
  \BibitemOpen
  \bibfield  {author} {\bibinfo {author} {\bibfnamefont {Seoungsu}\
  \bibnamefont {Lee}}, \bibinfo {author} {\bibfnamefont {W.}~\bibnamefont
  {Ratcliff}}, \bibinfo {author} {\bibfnamefont {S-W.}\ \bibnamefont {Cheong}},
  \ and\ \bibinfo {author} {\bibfnamefont {V.}~\bibnamefont {Kiryukhin}},\
  }\bibfield  {title} {\enquote {\bibinfo {title} {Electric field control of
  the magnetic state in {BiFeO$_3$} single crystals},}\ }\href
  {http://scitation.aip.org/content/aip/journal/apl/92/19/10.1063/1.2930678}
  {\bibfield  {journal} {\bibinfo  {journal} {Appl. Phys. Lett.}\ }\textbf
  {\bibinfo {volume} {92}},\ \bibinfo {eid} {192906} (\bibinfo {year}
  {2008}{\natexlab{a}})}\BibitemShut {NoStop}%
\bibitem [{\citenamefont {Catalan}\ and\ \citenamefont
  {Scott}(2009)}]{Catalan2009}%
  \BibitemOpen
  \bibfield  {author} {\bibinfo {author} {\bibfnamefont {Gustau}\ \bibnamefont
  {Catalan}}\ and\ \bibinfo {author} {\bibfnamefont {James~F.}\ \bibnamefont
  {Scott}},\ }\bibfield  {title} {\enquote {\bibinfo {title} {Physics and
  applications of bismuth ferrite},}\ }\href {\doibase 10.1002/adma.200802849}
  {\bibfield  {journal} {\bibinfo  {journal} {Advanced Materials}\ }\textbf
  {\bibinfo {volume} {21}},\ \bibinfo {pages} {2463} (\bibinfo {year}
  {2009})}\BibitemShut {NoStop}%
\bibitem [{\citenamefont {Spalding}\ and\ \citenamefont
  {Fiebig}(2005)}]{Spalding2005}%
  \BibitemOpen
  \bibfield  {author} {\bibinfo {author} {\bibfnamefont {Nicola~A.}\
  \bibnamefont {Spalding}}\ and\ \bibinfo {author} {\bibfnamefont {Manfred}\
  \bibnamefont {Fiebig}},\ }\bibfield  {title} {\enquote {\bibinfo {title} {The
  renaissance of magnetoelectric multiferroics},}\ }\href {\doibase
  10.1126/science.1113357} {\bibfield  {journal} {\bibinfo  {journal}
  {Science}\ }\textbf {\bibinfo {volume} {309}},\ \bibinfo {pages} {391--392}
  (\bibinfo {year} {2005})}\BibitemShut {NoStop}%
\bibitem [{\citenamefont {Cheong}\ and\ \citenamefont
  {Mostovoy}(2007)}]{Cheong2007}%
  \BibitemOpen
  \bibfield  {author} {\bibinfo {author} {\bibfnamefont {Sang-Wook}\
  \bibnamefont {Cheong}}\ and\ \bibinfo {author} {\bibfnamefont {Maxim}\
  \bibnamefont {Mostovoy}},\ }\bibfield  {title} {\enquote {\bibinfo {title}
  {Multiferroics: a magnetic twist for ferroelectricity},}\ }\href {\doibase
  10.1038/nmat1804} {\bibfield  {journal} {\bibinfo  {journal} {Nature Mater.}\
  }\textbf {\bibinfo {volume} {6}},\ \bibinfo {pages} {13--20} (\bibinfo {year}
  {2007})}\BibitemShut {NoStop}%
\bibitem [{\citenamefont {Senff}\ \emph {et~al.}(2007)\citenamefont {Senff},
  \citenamefont {Link}, \citenamefont {Hradil}, \citenamefont {Hiess},
  \citenamefont {Regnault}, \citenamefont {Sidis}, \citenamefont {Aliouane},
  \citenamefont {Argyriou},\ and\ \citenamefont {Braden}}]{senff2007}%
  \BibitemOpen
  \bibfield  {author} {\bibinfo {author} {\bibfnamefont {D.}~\bibnamefont
  {Senff}}, \bibinfo {author} {\bibfnamefont {P.}~\bibnamefont {Link}},
  \bibinfo {author} {\bibfnamefont {K.}~\bibnamefont {Hradil}}, \bibinfo
  {author} {\bibfnamefont {A.}~\bibnamefont {Hiess}}, \bibinfo {author}
  {\bibfnamefont {L.~P.}\ \bibnamefont {Regnault}}, \bibinfo {author}
  {\bibfnamefont {Y.}~\bibnamefont {Sidis}}, \bibinfo {author} {\bibfnamefont
  {N.}~\bibnamefont {Aliouane}}, \bibinfo {author} {\bibfnamefont {D.~N.}\
  \bibnamefont {Argyriou}}, \ and\ \bibinfo {author} {\bibfnamefont
  {M.}~\bibnamefont {Braden}},\ }\bibfield  {title} {\enquote {\bibinfo {title}
  {Magnetic excitations in multiferroic {TbMnO$_{3}$}: Evidence for a
  hybridized soft mode},}\ }\href {\doibase 10.1103/PhysRevLett.98.137206}
  {\bibfield  {journal} {\bibinfo  {journal} {Phys. Rev. Lett.}\ }\textbf
  {\bibinfo {volume} {98}},\ \bibinfo {pages} {137206} (\bibinfo {year}
  {2007})}\BibitemShut {NoStop}%
\bibitem [{\citenamefont {Vajk}\ \emph {et~al.}(2005)\citenamefont {Vajk},
  \citenamefont {Kenzelmann}, \citenamefont {Lynn}, \citenamefont {Kim},\ and\
  \citenamefont {Cheong}}]{vajk05}%
  \BibitemOpen
  \bibfield  {author} {\bibinfo {author} {\bibfnamefont {O.~P.}\ \bibnamefont
  {Vajk}}, \bibinfo {author} {\bibfnamefont {M.}~\bibnamefont {Kenzelmann}},
  \bibinfo {author} {\bibfnamefont {J.~W.}\ \bibnamefont {Lynn}}, \bibinfo
  {author} {\bibfnamefont {S.~B.}\ \bibnamefont {Kim}}, \ and\ \bibinfo
  {author} {\bibfnamefont {S.-W.}\ \bibnamefont {Cheong}},\ }\bibfield  {title}
  {\enquote {\bibinfo {title} {Magnetic order and spin dynamics in
  ferroelectric {HoMnO$_{3}$}},}\ }\href {\doibase
  10.1103/PhysRevLett.94.087601} {\bibfield  {journal} {\bibinfo  {journal}
  {Phys. Rev. Lett.}\ }\textbf {\bibinfo {volume} {94}},\ \bibinfo {pages}
  {087601} (\bibinfo {year} {2005})}\BibitemShut {NoStop}%
\bibitem [{\citenamefont {Jeong}\ \emph {et~al.}(2012)\citenamefont {Jeong},
  \citenamefont {Goremychkin}, \citenamefont {Guidi}, \citenamefont {Nakajima},
  \citenamefont {Jeon}, \citenamefont {Kim}, \citenamefont {Furukawa},
  \citenamefont {Kim}, \citenamefont {Lee}, \citenamefont {Kiryukhin},
  \citenamefont {Cheong},\ and\ \citenamefont {Park}}]{jeong12}%
  \BibitemOpen
  \bibfield  {author} {\bibinfo {author} {\bibfnamefont {Jaehong}\ \bibnamefont
  {Jeong}}, \bibinfo {author} {\bibfnamefont {E.~A.}\ \bibnamefont
  {Goremychkin}}, \bibinfo {author} {\bibfnamefont {T.}~\bibnamefont {Guidi}},
  \bibinfo {author} {\bibfnamefont {K.}~\bibnamefont {Nakajima}}, \bibinfo
  {author} {\bibfnamefont {Gun~Sang}\ \bibnamefont {Jeon}}, \bibinfo {author}
  {\bibfnamefont {Shin-Ae}\ \bibnamefont {Kim}}, \bibinfo {author}
  {\bibfnamefont {S.}~\bibnamefont {Furukawa}}, \bibinfo {author}
  {\bibfnamefont {Yong~Baek}\ \bibnamefont {Kim}}, \bibinfo {author}
  {\bibfnamefont {Seongsu}\ \bibnamefont {Lee}}, \bibinfo {author}
  {\bibfnamefont {V.}~\bibnamefont {Kiryukhin}}, \bibinfo {author}
  {\bibfnamefont {S-W.}\ \bibnamefont {Cheong}}, \ and\ \bibinfo {author}
  {\bibfnamefont {Je-Geun}\ \bibnamefont {Park}},\ }\bibfield  {title}
  {\enquote {\bibinfo {title} {Spin wave measurements over the full brillouin
  zone of multiferroic {BiFeO}$_{3}$},}\ }\href {\doibase
  10.1103/PhysRevLett.108.077202} {\bibfield  {journal} {\bibinfo  {journal}
  {Phys. Rev. Lett.}\ }\textbf {\bibinfo {volume} {108}},\ \bibinfo {pages}
  {077202} (\bibinfo {year} {2012})}\BibitemShut {NoStop}%
\bibitem [{\citenamefont {Matsuda}\ \emph {et~al.}(2012)\citenamefont
  {Matsuda}, \citenamefont {Fishman}, \citenamefont {Hong}, \citenamefont
  {Lee}, \citenamefont {Ushiyama}, \citenamefont {Yanagisawa}, \citenamefont
  {Tomioka},\ and\ \citenamefont {Ito}}]{matsuda12}%
  \BibitemOpen
  \bibfield  {author} {\bibinfo {author} {\bibfnamefont {M.}~\bibnamefont
  {Matsuda}}, \bibinfo {author} {\bibfnamefont {R.~S.}\ \bibnamefont
  {Fishman}}, \bibinfo {author} {\bibfnamefont {T.}~\bibnamefont {Hong}},
  \bibinfo {author} {\bibfnamefont {C.~H.}\ \bibnamefont {Lee}}, \bibinfo
  {author} {\bibfnamefont {T.}~\bibnamefont {Ushiyama}}, \bibinfo {author}
  {\bibfnamefont {Y.}~\bibnamefont {Yanagisawa}}, \bibinfo {author}
  {\bibfnamefont {Y.}~\bibnamefont {Tomioka}}, \ and\ \bibinfo {author}
  {\bibfnamefont {T.}~\bibnamefont {Ito}},\ }\bibfield  {title} {\enquote
  {\bibinfo {title} {Magnetic dispersion and anisotropy in multiferroic
  {BiFeO}$_{3}$},}\ }\href {\doibase 10.1103/PhysRevLett.109.067205} {\bibfield
   {journal} {\bibinfo  {journal} {Phys. Rev. Lett.}\ }\textbf {\bibinfo
  {volume} {109}},\ \bibinfo {pages} {067205} (\bibinfo {year}
  {2012})}\BibitemShut {NoStop}%
\bibitem [{\citenamefont {Jeong}\ \emph {et~al.}(2014)\citenamefont {Jeong},
  \citenamefont {Le}, \citenamefont {Bourges}, \citenamefont {Petit},
  \citenamefont {Furukawa}, \citenamefont {Kim}, \citenamefont {Lee},
  \citenamefont {Cheong},\ and\ \citenamefont {Park}}]{jeong14}%
  \BibitemOpen
  \bibfield  {author} {\bibinfo {author} {\bibfnamefont {Jaehong}\ \bibnamefont
  {Jeong}}, \bibinfo {author} {\bibfnamefont {Manh~Duc}\ \bibnamefont {Le}},
  \bibinfo {author} {\bibfnamefont {P.}~\bibnamefont {Bourges}}, \bibinfo
  {author} {\bibfnamefont {S.}~\bibnamefont {Petit}}, \bibinfo {author}
  {\bibfnamefont {S.}~\bibnamefont {Furukawa}}, \bibinfo {author}
  {\bibfnamefont {Shin-Ae}\ \bibnamefont {Kim}}, \bibinfo {author}
  {\bibfnamefont {Seongsu}\ \bibnamefont {Lee}}, \bibinfo {author}
  {\bibfnamefont {S-W.}\ \bibnamefont {Cheong}}, \ and\ \bibinfo {author}
  {\bibfnamefont {Je-Geun}\ \bibnamefont {Park}},\ }\bibfield  {title}
  {\enquote {\bibinfo {title} {Temperature-dependent interplay of
  {D}zyaloshinskii-{M}oriya interaction and single-ion anisotropy in
  multiferroic {BiFeO}$_{3}$},}\ }\href {\doibase
  10.1103/PhysRevLett.113.107202} {\bibfield  {journal} {\bibinfo  {journal}
  {Phys. Rev. Lett.}\ }\textbf {\bibinfo {volume} {113}},\ \bibinfo {pages}
  {107202} (\bibinfo {year} {2014})}\BibitemShut {NoStop}%
\bibitem [{\citenamefont {Park}\ \emph {et~al.}(2016)\citenamefont {Park},
  \citenamefont {Oh}, \citenamefont {Leiner}, \citenamefont {Jeong},
  \citenamefont {Rule}, \citenamefont {Le},\ and\ \citenamefont
  {Park}}]{park2016}%
  \BibitemOpen
  \bibfield  {author} {\bibinfo {author} {\bibfnamefont {Kisoo}\ \bibnamefont
  {Park}}, \bibinfo {author} {\bibfnamefont {Joosung}\ \bibnamefont {Oh}},
  \bibinfo {author} {\bibfnamefont {Jonathan~C.}\ \bibnamefont {Leiner}},
  \bibinfo {author} {\bibfnamefont {Jaehong}\ \bibnamefont {Jeong}}, \bibinfo
  {author} {\bibfnamefont {Kirrily~C.}\ \bibnamefont {Rule}}, \bibinfo {author}
  {\bibfnamefont {Manh~Duc}\ \bibnamefont {Le}}, \ and\ \bibinfo {author}
  {\bibfnamefont {Je-Geun}\ \bibnamefont {Park}},\ }\bibfield  {title}
  {\enquote {\bibinfo {title} {Magnon-phonon coupling and two-magnon continuum
  in the two-dimensional triangular antiferromagnet {CuCrO}$_{2}$},}\ }\href
  {\doibase 10.1103/PhysRevB.94.104421} {\bibfield  {journal} {\bibinfo
  {journal} {Phys. Rev. B}\ }\textbf {\bibinfo {volume} {94}},\ \bibinfo
  {pages} {104421} (\bibinfo {year} {2016})}\BibitemShut {NoStop}%
\bibitem [{\citenamefont {Schneeloch}\ \emph {et~al.}(2015)\citenamefont
  {Schneeloch}, \citenamefont {Xu}, \citenamefont {Wen}, \citenamefont
  {Gehring}, \citenamefont {Stock}, \citenamefont {Matsuda}, \citenamefont
  {Winn}, \citenamefont {Gu}, \citenamefont {Shapiro}, \citenamefont
  {Birgeneau}, \citenamefont {Ushiyama}, \citenamefont {Yanagisawa},
  \citenamefont {Tomioka}, \citenamefont {Ito},\ and\ \citenamefont
  {Xu}}]{schneeloch2015}%
  \BibitemOpen
  \bibfield  {author} {\bibinfo {author} {\bibfnamefont {John~A.}\ \bibnamefont
  {Schneeloch}}, \bibinfo {author} {\bibfnamefont {Zhijun}\ \bibnamefont {Xu}},
  \bibinfo {author} {\bibfnamefont {Jinsheng}\ \bibnamefont {Wen}}, \bibinfo
  {author} {\bibfnamefont {P.~M.}\ \bibnamefont {Gehring}}, \bibinfo {author}
  {\bibfnamefont {C.}~\bibnamefont {Stock}}, \bibinfo {author} {\bibfnamefont
  {M.}~\bibnamefont {Matsuda}}, \bibinfo {author} {\bibfnamefont
  {B.}~\bibnamefont {Winn}}, \bibinfo {author} {\bibfnamefont {Genda}\
  \bibnamefont {Gu}}, \bibinfo {author} {\bibfnamefont {Stephen~M.}\
  \bibnamefont {Shapiro}}, \bibinfo {author} {\bibfnamefont {R.~J.}\
  \bibnamefont {Birgeneau}}, \bibinfo {author} {\bibfnamefont {T.}~\bibnamefont
  {Ushiyama}}, \bibinfo {author} {\bibfnamefont {Y.}~\bibnamefont
  {Yanagisawa}}, \bibinfo {author} {\bibfnamefont {Y.}~\bibnamefont {Tomioka}},
  \bibinfo {author} {\bibfnamefont {T.}~\bibnamefont {Ito}}, \ and\ \bibinfo
  {author} {\bibfnamefont {Guangyong}\ \bibnamefont {Xu}},\ }\bibfield  {title}
  {\enquote {\bibinfo {title} {Neutron inelastic scattering measurements of
  low-energy phonons in the multiferroic {BiFeO}$_{3}$},}\ }\href {\doibase
  10.1103/PhysRevB.91.064301} {\bibfield  {journal} {\bibinfo  {journal} {Phys.
  Rev. B}\ }\textbf {\bibinfo {volume} {91}},\ \bibinfo {pages} {064301}
  (\bibinfo {year} {2015})}\BibitemShut {NoStop}%
\bibitem [{\citenamefont {Petit}\ \emph {et~al.}(2007)\citenamefont {Petit},
  \citenamefont {Moussa}, \citenamefont {Hennion}, \citenamefont {Pailh\`es},
  \citenamefont {Pinsard-Gaudart},\ and\ \citenamefont {Ivanov}}]{Petit2007}%
  \BibitemOpen
  \bibfield  {author} {\bibinfo {author} {\bibfnamefont {S.}~\bibnamefont
  {Petit}}, \bibinfo {author} {\bibfnamefont {F.}~\bibnamefont {Moussa}},
  \bibinfo {author} {\bibfnamefont {M.}~\bibnamefont {Hennion}}, \bibinfo
  {author} {\bibfnamefont {S.}~\bibnamefont {Pailh\`es}}, \bibinfo {author}
  {\bibfnamefont {L.}~\bibnamefont {Pinsard-Gaudart}}, \ and\ \bibinfo {author}
  {\bibfnamefont {A.}~\bibnamefont {Ivanov}},\ }\bibfield  {title} {\enquote
  {\bibinfo {title} {Spin phonon coupling in hexagonal multiferroic
  {YMnO}$_{3}$},}\ }\href {\doibase 10.1103/PhysRevLett.99.266604} {\bibfield
  {journal} {\bibinfo  {journal} {Phys. Rev. Lett.}\ }\textbf {\bibinfo
  {volume} {99}},\ \bibinfo {pages} {266604} (\bibinfo {year}
  {2007})}\BibitemShut {NoStop}%
\bibitem [{\citenamefont {Sim}\ \emph {et~al.}(2016)\citenamefont {Sim},
  \citenamefont {Oh}, \citenamefont {Jeong}, \citenamefont {Le},\ and\
  \citenamefont {Park}}]{Sim2016}%
  \BibitemOpen
  \bibfield  {author} {\bibinfo {author} {\bibfnamefont {Hasung}\ \bibnamefont
  {Sim}}, \bibinfo {author} {\bibfnamefont {Joosung}\ \bibnamefont {Oh}},
  \bibinfo {author} {\bibfnamefont {Jaehong}\ \bibnamefont {Jeong}}, \bibinfo
  {author} {\bibfnamefont {Manh~Duc}\ \bibnamefont {Le}}, \ and\ \bibinfo
  {author} {\bibfnamefont {Je-Geun}\ \bibnamefont {Park}},\ }\bibfield  {title}
  {\enquote {\bibinfo {title} {{Hexagonal {\it R}MnO${_3}$: a model system for
  two-dimensional triangular lattice antiferromagnets}},}\ }\href {\doibase
  10.1107/S2052520615022106} {\bibfield  {journal} {\bibinfo  {journal} {Acta
  Cryst. B}\ }\textbf {\bibinfo {volume} {72}},\ \bibinfo {pages} {3--19}
  (\bibinfo {year} {2016})}\BibitemShut {NoStop}%
\bibitem [{\citenamefont {Roessli}\ \emph {et~al.}(2005)\citenamefont
  {Roessli}, \citenamefont {Gvasaliya}, \citenamefont {Pomjakushina},\ and\
  \citenamefont {Conder}}]{Roessli2005}%
  \BibitemOpen
  \bibfield  {author} {\bibinfo {author} {\bibfnamefont {B.}~\bibnamefont
  {Roessli}}, \bibinfo {author} {\bibfnamefont {S.~N.}\ \bibnamefont
  {Gvasaliya}}, \bibinfo {author} {\bibfnamefont {E.}~\bibnamefont
  {Pomjakushina}}, \ and\ \bibinfo {author} {\bibfnamefont {K.}~\bibnamefont
  {Conder}},\ }\bibfield  {title} {\enquote {\bibinfo {title} {Spin
  fluctuations in the stacked-triangular antiferromagnet {YMnO}$_3$},}\ }\href
  {\doibase 10.1134/1.1931017} {\bibfield  {journal} {\bibinfo  {journal} {J.
  Exp. Theor. Phys. Lett.}\ }\textbf {\bibinfo {volume} {81}},\ \bibinfo
  {pages} {287--291} (\bibinfo {year} {2005})}\BibitemShut {NoStop}%
\bibitem [{\citenamefont {Lee}\ \emph {et~al.}(2008{\natexlab{b}})\citenamefont
  {Lee}, \citenamefont {Pirogov}, \citenamefont {Kang}, \citenamefont {Jang},
  \citenamefont {Yonemura}, \citenamefont {Kamiyama}, \citenamefont {Cheong},
  \citenamefont {Gozzo}, \citenamefont {Shin}, \citenamefont {Kimura},
  \citenamefont {Noda},\ and\ \citenamefont {Park}}]{Lee2008}%
  \BibitemOpen
  \bibfield  {author} {\bibinfo {author} {\bibfnamefont {Seongsu}\ \bibnamefont
  {Lee}}, \bibinfo {author} {\bibfnamefont {A.}~\bibnamefont {Pirogov}},
  \bibinfo {author} {\bibfnamefont {Misun}\ \bibnamefont {Kang}}, \bibinfo
  {author} {\bibfnamefont {Kwang-Hyun}\ \bibnamefont {Jang}}, \bibinfo {author}
  {\bibfnamefont {M.}~\bibnamefont {Yonemura}}, \bibinfo {author}
  {\bibfnamefont {T.}~\bibnamefont {Kamiyama}}, \bibinfo {author}
  {\bibfnamefont {S.-W.}\ \bibnamefont {Cheong}}, \bibinfo {author}
  {\bibfnamefont {F.}~\bibnamefont {Gozzo}}, \bibinfo {author} {\bibfnamefont
  {Namsoo}\ \bibnamefont {Shin}}, \bibinfo {author} {\bibfnamefont
  {H.}~\bibnamefont {Kimura}}, \bibinfo {author} {\bibfnamefont
  {Y.}~\bibnamefont {Noda}}, \ and\ \bibinfo {author} {\bibfnamefont {J.-G.}\
  \bibnamefont {Park}},\ }\bibfield  {title} {\enquote {\bibinfo {title} {Giant
  magneto-elastic coupling in multiferroic hexagonal manganites},}\ }\href
  {\doibase 10.1038/nature06507} {\bibfield  {journal} {\bibinfo  {journal}
  {Nature (London)}\ }\textbf {\bibinfo {volume} {451}},\ \bibinfo {pages}
  {805--808} (\bibinfo {year} {2008}{\natexlab{b}})}\BibitemShut {NoStop}%
\bibitem [{\citenamefont {Varignon}\ \emph {et~al.}(2013)\citenamefont
  {Varignon}, \citenamefont {Petit}, \citenamefont {Gell\'e},\ and\
  \citenamefont {Lepetit}}]{varignon2013}%
  \BibitemOpen
  \bibfield  {author} {\bibinfo {author} {\bibfnamefont {J.}~\bibnamefont
  {Varignon}}, \bibinfo {author} {\bibfnamefont {S.}~\bibnamefont {Petit}},
  \bibinfo {author} {\bibfnamefont {A.}~\bibnamefont {Gell\'e}}, \ and\
  \bibinfo {author} {\bibfnamefont {M.~B.}\ \bibnamefont {Lepetit}},\
  }\bibfield  {title} {\enquote {\bibinfo {title} {An ab initio study of
  magneto-electric coupling of {YMnO}$_3$},}\ }\href
  {http://stacks.iop.org/0953-8984/25/i=49/a=496004} {\bibfield  {journal}
  {\bibinfo  {journal} {Journal of Physics: Condensed Matter}\ }\textbf
  {\bibinfo {volume} {25}},\ \bibinfo {pages} {496004} (\bibinfo {year}
  {2013})}\BibitemShut {NoStop}%
\bibitem [{\citenamefont {Chatterji}\ \emph {et~al.}(2012)\citenamefont
  {Chatterji}, \citenamefont {Ouladdiaf}, \citenamefont {Henry},\ and\
  \citenamefont {Bhattacharya}}]{Chatterji2012}%
  \BibitemOpen
  \bibfield  {author} {\bibinfo {author} {\bibfnamefont {Tapan}\ \bibnamefont
  {Chatterji}}, \bibinfo {author} {\bibfnamefont {Bachir}\ \bibnamefont
  {Ouladdiaf}}, \bibinfo {author} {\bibfnamefont {Paul~F}\ \bibnamefont
  {Henry}}, \ and\ \bibinfo {author} {\bibfnamefont {Dipten}\ \bibnamefont
  {Bhattacharya}},\ }\bibfield  {title} {\enquote {\bibinfo {title}
  {Magnetoelastic effects in multiferroic {YMnO}$_3$},}\ }\href
  {http://stacks.iop.org/0953-8984/24/i=33/a=336003} {\bibfield  {journal}
  {\bibinfo  {journal} {J. Phys.: Cond. Matt.}\ }\textbf {\bibinfo {volume}
  {24}},\ \bibinfo {pages} {336003} (\bibinfo {year} {2012})}\BibitemShut
  {NoStop}%
\bibitem [{\citenamefont {Howard}\ \emph {et~al.}(2013)\citenamefont {Howard},
  \citenamefont {Campbell}, \citenamefont {Stokes}, \citenamefont {Carpenter},\
  and\ \citenamefont {Thomson}}]{Howard2013}%
  \BibitemOpen
  \bibfield  {author} {\bibinfo {author} {\bibfnamefont {Christopher~J.}\
  \bibnamefont {Howard}}, \bibinfo {author} {\bibfnamefont {Branton~J.}\
  \bibnamefont {Campbell}}, \bibinfo {author} {\bibfnamefont {Harold~T.}\
  \bibnamefont {Stokes}}, \bibinfo {author} {\bibfnamefont {Michael~A.}\
  \bibnamefont {Carpenter}}, \ and\ \bibinfo {author} {\bibfnamefont
  {Richard~I.}\ \bibnamefont {Thomson}},\ }\bibfield  {title} {\enquote
  {\bibinfo {title} {{Crystal and magnetic structures of hexagonal
  YMnO$_3$}},}\ }\href {\doibase 10.1107/S205251921302993X} {\bibfield
  {journal} {\bibinfo  {journal} {Acta Cryst. B}\ }\textbf {\bibinfo {volume}
  {69}},\ \bibinfo {pages} {534--540} (\bibinfo {year} {2013})}\BibitemShut
  {NoStop}%
\bibitem [{\citenamefont {Thomson}\ \emph {et~al.}(2014)\citenamefont
  {Thomson}, \citenamefont {Chatterji}, \citenamefont {Howard}, \citenamefont
  {Palstra},\ and\ \citenamefont {Carpenter}}]{thomson14}%
  \BibitemOpen
  \bibfield  {author} {\bibinfo {author} {\bibfnamefont {R~I}\ \bibnamefont
  {Thomson}}, \bibinfo {author} {\bibfnamefont {T}~\bibnamefont {Chatterji}},
  \bibinfo {author} {\bibfnamefont {C~J}\ \bibnamefont {Howard}}, \bibinfo
  {author} {\bibfnamefont {T~T~M}\ \bibnamefont {Palstra}}, \ and\ \bibinfo
  {author} {\bibfnamefont {M~A}\ \bibnamefont {Carpenter}},\ }\bibfield
  {title} {\enquote {\bibinfo {title} {Elastic anomalies associated with
  structural and magnetic phase transitions in single crystal hexagonal
  {YMnO$_3$}},}\ }\href {http://stacks.iop.org/0953-8984/26/i=4/a=045901}
  {\bibfield  {journal} {\bibinfo  {journal} {J. Phys.: Cond. Matt.}\ }\textbf
  {\bibinfo {volume} {26}},\ \bibinfo {pages} {045901} (\bibinfo {year}
  {2014})}\BibitemShut {NoStop}%
\bibitem [{\citenamefont {Brown}\ and\ \citenamefont
  {Chatterji}(2006)}]{Brown2006}%
  \BibitemOpen
  \bibfield  {author} {\bibinfo {author} {\bibfnamefont {P~J}\ \bibnamefont
  {Brown}}\ and\ \bibinfo {author} {\bibfnamefont {T}~\bibnamefont
  {Chatterji}},\ }\bibfield  {title} {\enquote {\bibinfo {title} {Neutron
  diffraction and polarimetric study of the magnetic and crystal structures of
  {HoMnO$_3$} and {YMnO$_3$}},}\ }\href
  {http://stacks.iop.org/0953-8984/18/i=44/a=008} {\bibfield  {journal}
  {\bibinfo  {journal} {J. Phys.: Cond. Matt.}\ }\textbf {\bibinfo {volume}
  {18}},\ \bibinfo {pages} {10085} (\bibinfo {year} {2006})}\BibitemShut
  {NoStop}%
\bibitem [{\citenamefont {Demmel}\ and\ \citenamefont
  {Chatterji}(2007)}]{Chatterji2007}%
  \BibitemOpen
  \bibfield  {author} {\bibinfo {author} {\bibfnamefont {Franz}\ \bibnamefont
  {Demmel}}\ and\ \bibinfo {author} {\bibfnamefont {Tapan}\ \bibnamefont
  {Chatterji}},\ }\bibfield  {title} {\enquote {\bibinfo {title} {Persistent
  spin waves above the {N\'eel} temperature in {YMnO}$_{3}$},}\ }\href
  {\doibase 10.1103/PhysRevB.76.212402} {\bibfield  {journal} {\bibinfo
  {journal} {Phys. Rev. B}\ }\textbf {\bibinfo {volume} {76}},\ \bibinfo
  {pages} {212402} (\bibinfo {year} {2007})}\BibitemShut {NoStop}%
\bibitem [{\citenamefont {Singh}\ \emph {et~al.}(2013)\citenamefont {Singh},
  \citenamefont {Lepetit}, \citenamefont {Simon}, \citenamefont {Bellido},
  \citenamefont {Pailh\`es}, \citenamefont {Varignon},\ and\ \citenamefont
  {Muer}}]{singh2013}%
  \BibitemOpen
  \bibfield  {author} {\bibinfo {author} {\bibfnamefont {Kiran}\ \bibnamefont
  {Singh}}, \bibinfo {author} {\bibfnamefont {Marie-Bernadette}\ \bibnamefont
  {Lepetit}}, \bibinfo {author} {\bibfnamefont {Charles}\ \bibnamefont
  {Simon}}, \bibinfo {author} {\bibfnamefont {Natalia}\ \bibnamefont
  {Bellido}}, \bibinfo {author} {\bibfnamefont {St\'ephane}\ \bibnamefont
  {Pailh\`es}}, \bibinfo {author} {\bibfnamefont {Julien}\ \bibnamefont
  {Varignon}}, \ and\ \bibinfo {author} {\bibfnamefont {Albin~De}\ \bibnamefont
  {Muer}},\ }\bibfield  {title} {\enquote {\bibinfo {title} {Analysis of the
  multiferroicity in the hexagonal manganite {YMnO}$_3$},}\ }\href
  {http://stacks.iop.org/0953-8984/25/i=41/a=416002} {\bibfield  {journal}
  {\bibinfo  {journal} {J. Phys.: Cond. Matt.}\ }\textbf {\bibinfo {volume}
  {25}},\ \bibinfo {pages} {416002} (\bibinfo {year} {2013})}\BibitemShut
  {NoStop}%
\bibitem [{\citenamefont {Momma}\ and\ \citenamefont
  {Izumi}(2011)}]{Momma2011}%
  \BibitemOpen
  \bibfield  {author} {\bibinfo {author} {\bibfnamefont {Koichi}\ \bibnamefont
  {Momma}}\ and\ \bibinfo {author} {\bibfnamefont {Fujio}\ \bibnamefont
  {Izumi}},\ }\bibfield  {title} {\enquote {\bibinfo {title} {{{\it VESTA3} for
  three-dimensional visualization of crystal, volumetric and morphology
  data}},}\ }\href {\doibase 10.1107/S0021889811038970} {\bibfield  {journal}
  {\bibinfo  {journal} {J. Appl. Cryst.}\ }\textbf {\bibinfo {volume} {44}},\
  \bibinfo {pages} {1272--1276} (\bibinfo {year} {2011})}\BibitemShut {NoStop}%
\bibitem [{\citenamefont {Sato}\ \emph {et~al.}(2003)\citenamefont {Sato},
  \citenamefont {Lee}, \citenamefont {Katsufuji}, \citenamefont {Masaki},
  \citenamefont {Park}, \citenamefont {Copley},\ and\ \citenamefont
  {Takagi}}]{Sato2003}%
  \BibitemOpen
  \bibfield  {author} {\bibinfo {author} {\bibfnamefont {T.~J.}\ \bibnamefont
  {Sato}}, \bibinfo {author} {\bibfnamefont {S.-H.}\ \bibnamefont {Lee}},
  \bibinfo {author} {\bibfnamefont {T.}~\bibnamefont {Katsufuji}}, \bibinfo
  {author} {\bibfnamefont {M.}~\bibnamefont {Masaki}}, \bibinfo {author}
  {\bibfnamefont {S.}~\bibnamefont {Park}}, \bibinfo {author} {\bibfnamefont
  {J.~R.~D.}\ \bibnamefont {Copley}}, \ and\ \bibinfo {author} {\bibfnamefont
  {H.}~\bibnamefont {Takagi}},\ }\bibfield  {title} {\enquote {\bibinfo {title}
  {Unconventional spin fluctuations in the hexagonal antiferromagnet
  {YMnO}$_{3}$},}\ }\href {\doibase 10.1103/PhysRevB.68.014432} {\bibfield
  {journal} {\bibinfo  {journal} {Phys. Rev. B}\ }\textbf {\bibinfo {volume}
  {68}},\ \bibinfo {pages} {014432} (\bibinfo {year} {2003})}\BibitemShut
  {NoStop}%
\bibitem [{\citenamefont {Pailh\`es}\ \emph {et~al.}(2009)\citenamefont
  {Pailh\`es}, \citenamefont {Fabr\`eges}, \citenamefont {R\'egnault},
  \citenamefont {Pinsard-Godart}, \citenamefont {Mirebeau}, \citenamefont
  {Moussa}, \citenamefont {Hennion},\ and\ \citenamefont {Petit}}]{Pailhes09}%
  \BibitemOpen
  \bibfield  {author} {\bibinfo {author} {\bibfnamefont {S.}~\bibnamefont
  {Pailh\`es}}, \bibinfo {author} {\bibfnamefont {X.}~\bibnamefont
  {Fabr\`eges}}, \bibinfo {author} {\bibfnamefont {L.~P.}\ \bibnamefont
  {R\'egnault}}, \bibinfo {author} {\bibfnamefont {L.}~\bibnamefont
  {Pinsard-Godart}}, \bibinfo {author} {\bibfnamefont {I.}~\bibnamefont
  {Mirebeau}}, \bibinfo {author} {\bibfnamefont {F.}~\bibnamefont {Moussa}},
  \bibinfo {author} {\bibfnamefont {M.}~\bibnamefont {Hennion}}, \ and\
  \bibinfo {author} {\bibfnamefont {S.}~\bibnamefont {Petit}},\ }\bibfield
  {title} {\enquote {\bibinfo {title} {Hybrid goldstone modes in multiferroic
  {YMnO}$_3$ studied by polarized inelastic neutron scattering},}\ }\href
  {\doibase 10.1103/PhysRevB.79.134409} {\bibfield  {journal} {\bibinfo
  {journal} {Phys. Rev. B}\ }\textbf {\bibinfo {volume} {79}},\ \bibinfo
  {pages} {134409} (\bibinfo {year} {2009})}\BibitemShut {NoStop}%
\bibitem [{\citenamefont {Oh}\ \emph {et~al.}(2016)\citenamefont {Oh},
  \citenamefont {Le}, \citenamefont {Nahm}, \citenamefont {Sim}, \citenamefont
  {Jeong}, \citenamefont {Perring}, \citenamefont {Woo}, \citenamefont
  {Nakajima}, \citenamefont {Ohira-Kawamura}, \citenamefont {Yamani},
  \citenamefont {Yoshida}, \citenamefont {Eisaki}, \citenamefont {Cheong},
  \citenamefont {Chernyshev},\ and\ \citenamefont {Park}}]{Oh2016}%
  \BibitemOpen
  \bibfield  {author} {\bibinfo {author} {\bibfnamefont {Joosung}\ \bibnamefont
  {Oh}}, \bibinfo {author} {\bibfnamefont {Manh~Duc}\ \bibnamefont {Le}},
  \bibinfo {author} {\bibfnamefont {Ho-Hyun}\ \bibnamefont {Nahm}}, \bibinfo
  {author} {\bibfnamefont {Hasung}\ \bibnamefont {Sim}}, \bibinfo {author}
  {\bibfnamefont {Jaehong}\ \bibnamefont {Jeong}}, \bibinfo {author}
  {\bibfnamefont {T.~G.}\ \bibnamefont {Perring}}, \bibinfo {author}
  {\bibfnamefont {Hyungje}\ \bibnamefont {Woo}}, \bibinfo {author}
  {\bibfnamefont {Kenji}\ \bibnamefont {Nakajima}}, \bibinfo {author}
  {\bibfnamefont {Seiko}\ \bibnamefont {Ohira-Kawamura}}, \bibinfo {author}
  {\bibfnamefont {Zahra}\ \bibnamefont {Yamani}}, \bibinfo {author}
  {\bibfnamefont {Y.}~\bibnamefont {Yoshida}}, \bibinfo {author} {\bibfnamefont
  {H.}~\bibnamefont {Eisaki}}, \bibinfo {author} {\bibfnamefont {S.~W.}\
  \bibnamefont {Cheong}}, \bibinfo {author} {\bibfnamefont {A.~L.}\
  \bibnamefont {Chernyshev}}, \ and\ \bibinfo {author} {\bibfnamefont
  {Je-Geun}\ \bibnamefont {Park}},\ }\bibfield  {title} {\enquote {\bibinfo
  {title} {Spontaneous decays of magneto-elastic excitations in non-collinear
  antiferromagnet {(Y,Lu)MnO}$_3$},}\ }\href
  {http://dx.doi.org/10.1038/ncomms13146} {\bibfield  {journal} {\bibinfo
  {journal} {Nat. Commun.}\ }\textbf {\bibinfo {volume} {7}},\ \bibinfo {pages}
  {13146} (\bibinfo {year} {2016})}\BibitemShut {NoStop}%
\bibitem [{\citenamefont {Lancaster}\ \emph {et~al.}(2007)\citenamefont
  {Lancaster}, \citenamefont {Blundell}, \citenamefont {Andreica},
  \citenamefont {Janoschek}, \citenamefont {Roessli}, \citenamefont
  {Gvasaliya}, \citenamefont {Conder}, \citenamefont {Pomjakushina},
  \citenamefont {Brooks}, \citenamefont {Baker}, \citenamefont {Prabhakaran},
  \citenamefont {Hayes},\ and\ \citenamefont {Pratt}}]{Lancaster2007}%
  \BibitemOpen
  \bibfield  {author} {\bibinfo {author} {\bibfnamefont {T.}~\bibnamefont
  {Lancaster}}, \bibinfo {author} {\bibfnamefont {S.~J.}\ \bibnamefont
  {Blundell}}, \bibinfo {author} {\bibfnamefont {D.}~\bibnamefont {Andreica}},
  \bibinfo {author} {\bibfnamefont {M.}~\bibnamefont {Janoschek}}, \bibinfo
  {author} {\bibfnamefont {B.}~\bibnamefont {Roessli}}, \bibinfo {author}
  {\bibfnamefont {S.~N.}\ \bibnamefont {Gvasaliya}}, \bibinfo {author}
  {\bibfnamefont {K.}~\bibnamefont {Conder}}, \bibinfo {author} {\bibfnamefont
  {E.}~\bibnamefont {Pomjakushina}}, \bibinfo {author} {\bibfnamefont {M.~L.}\
  \bibnamefont {Brooks}}, \bibinfo {author} {\bibfnamefont {P.~J.}\
  \bibnamefont {Baker}}, \bibinfo {author} {\bibfnamefont {D.}~\bibnamefont
  {Prabhakaran}}, \bibinfo {author} {\bibfnamefont {W.}~\bibnamefont {Hayes}},
  \ and\ \bibinfo {author} {\bibfnamefont {F.~L.}\ \bibnamefont {Pratt}},\
  }\bibfield  {title} {\enquote {\bibinfo {title} {Magnetism in geometrically
  frustrated {YMnO}$_{3}$ under hydrostatic pressure studied with muon spin
  relaxation},}\ }\href {\doibase 10.1103/PhysRevLett.98.197203} {\bibfield
  {journal} {\bibinfo  {journal} {Phys. Rev. Lett.}\ }\textbf {\bibinfo
  {volume} {98}},\ \bibinfo {pages} {197203} (\bibinfo {year}
  {2007})}\BibitemShut {NoStop}%
\bibitem [{\citenamefont {Bahl}\ \emph {et~al.}(2004)\citenamefont {Bahl},
  \citenamefont {Andersen}, \citenamefont {Klausen},\ and\ \citenamefont
  {Lefmann}}]{bahlrita1}%
  \BibitemOpen
  \bibfield  {author} {\bibinfo {author} {\bibfnamefont {C.R.H.}\ \bibnamefont
  {Bahl}}, \bibinfo {author} {\bibfnamefont {P.}~\bibnamefont {Andersen}},
  \bibinfo {author} {\bibfnamefont {S.N.}\ \bibnamefont {Klausen}}, \ and\
  \bibinfo {author} {\bibfnamefont {K.}~\bibnamefont {Lefmann}},\ }\bibfield
  {title} {\enquote {\bibinfo {title} {The monochromatic imaging mode of a
  {RITA}-type neutron spectrometer},}\ }\href {\doibase
  http://dx.doi.org/10.1016/j.nimb.2004.07.005} {\bibfield  {journal} {\bibinfo
   {journal} {Nucl. Instr. Meth. B}\ }\textbf {\bibinfo {volume} {226}},\
  \bibinfo {pages} {667 -- 681} (\bibinfo {year} {2004})}\BibitemShut {NoStop}%
\bibitem [{\citenamefont {Bahl}\ \emph {et~al.}(2006)\citenamefont {Bahl},
  \citenamefont {Lefmann}, \citenamefont {Abrahamsen}, \citenamefont
  {R{}\o{}nnow}, \citenamefont {Saxild}, \citenamefont {Jensen}, \citenamefont
  {Udby}, \citenamefont {Andersen}, \citenamefont {Christensen}, \citenamefont
  {Jakobsen}, \citenamefont {Larsen}, \citenamefont {H\"afliger}, \citenamefont
  {Streule},\ and\ \citenamefont {Niedermayer}}]{bahlrita2}%
  \BibitemOpen
  \bibfield  {author} {\bibinfo {author} {\bibfnamefont {C.R.H.}\ \bibnamefont
  {Bahl}}, \bibinfo {author} {\bibfnamefont {K.}~\bibnamefont {Lefmann}},
  \bibinfo {author} {\bibfnamefont {A.B.}\ \bibnamefont {Abrahamsen}}, \bibinfo
  {author} {\bibfnamefont {H.M.}\ \bibnamefont {R{}\o{}nnow}}, \bibinfo
  {author} {\bibfnamefont {F.}~\bibnamefont {Saxild}}, \bibinfo {author}
  {\bibfnamefont {T.B.S.}\ \bibnamefont {Jensen}}, \bibinfo {author}
  {\bibfnamefont {L.}~\bibnamefont {Udby}}, \bibinfo {author} {\bibfnamefont
  {N.H.}\ \bibnamefont {Andersen}}, \bibinfo {author} {\bibfnamefont {N.B.}\
  \bibnamefont {Christensen}}, \bibinfo {author} {\bibfnamefont {H.S.}\
  \bibnamefont {Jakobsen}}, \bibinfo {author} {\bibfnamefont {T.}~\bibnamefont
  {Larsen}}, \bibinfo {author} {\bibfnamefont {P.S.}\ \bibnamefont
  {H\"afliger}}, \bibinfo {author} {\bibfnamefont {S.}~\bibnamefont {Streule}},
  \ and\ \bibinfo {author} {\bibfnamefont {Ch.}\ \bibnamefont {Niedermayer}},\
  }\bibfield  {title} {\enquote {\bibinfo {title} {Inelastic neutron scattering
  experiments with the monochromatic imaging mode of the rita-ii
  spectrometer},}\ }\href {\doibase
  http://dx.doi.org/10.1016/j.nimb.2006.01.023} {\bibfield  {journal} {\bibinfo
   {journal} {Nuclear Instruments and Methods in Physics Research Section B:
  Beam Interactions with Materials and Atoms}\ }\textbf {\bibinfo {volume}
  {246}},\ \bibinfo {pages} {452 -- 462} (\bibinfo {year} {2006})}\BibitemShut
  {NoStop}%
\bibitem [{\citenamefont {Stuhr}\ \emph {et~al.}(2017)\citenamefont {Stuhr},
  \citenamefont {Roessli}, \citenamefont {Gvasaliya}, \citenamefont {nnow},
  \citenamefont {Filges}, \citenamefont {Graf}, \citenamefont {Bollhalder},
  \citenamefont {Hohl}, \citenamefont {B\"urge}, \citenamefont {Schild},
  \citenamefont {Holitzner}, \citenamefont {C.}, \citenamefont {Keller},\ and\
  \citenamefont {M\"uhlebach}}]{Stuhr2017}%
  \BibitemOpen
  \bibfield  {author} {\bibinfo {author} {\bibfnamefont {U.}~\bibnamefont
  {Stuhr}}, \bibinfo {author} {\bibfnamefont {B.}~\bibnamefont {Roessli}},
  \bibinfo {author} {\bibfnamefont {S.}~\bibnamefont {Gvasaliya}}, \bibinfo
  {author} {\bibfnamefont {H.M.~R\o}\ \bibnamefont {nnow}}, \bibinfo {author}
  {\bibfnamefont {U.}~\bibnamefont {Filges}}, \bibinfo {author} {\bibfnamefont
  {D.}~\bibnamefont {Graf}}, \bibinfo {author} {\bibfnamefont {A.}~\bibnamefont
  {Bollhalder}}, \bibinfo {author} {\bibfnamefont {D.}~\bibnamefont {Hohl}},
  \bibinfo {author} {\bibfnamefont {R.}~\bibnamefont {B\"urge}}, \bibinfo
  {author} {\bibfnamefont {M.}~\bibnamefont {Schild}}, \bibinfo {author}
  {\bibfnamefont {L.}~\bibnamefont {Holitzner}}, \bibinfo {author}
  {\bibfnamefont {Kaegi}\ \bibnamefont {C.}}, \bibinfo {author} {\bibfnamefont
  {P.}~\bibnamefont {Keller}}, \ and\ \bibinfo {author} {\bibfnamefont
  {T.}~\bibnamefont {M\"uhlebach}},\ }\bibfield  {title} {\enquote {\bibinfo
  {title} {The thermal triple-axis-spectrometer {EIGER} at the continuous
  spallation source {SINQ}},}\ }\href {\doibase
  https://doi.org/10.1016/j.nima.2017.02.003} {\bibfield  {journal} {\bibinfo
  {journal} {Nuclear Instruments and Methods in Physics Research Section A:
  Accelerators, Spectrometers, Detectors and Associated Equipment}\ }\textbf
  {\bibinfo {volume} {853}},\ \bibinfo {pages} {16 -- 19} (\bibinfo {year}
  {2017})}\BibitemShut {NoStop}%
\bibitem [{\citenamefont {Williams}(1988)}]{polarizationtextbook}%
  \BibitemOpen
  \bibfield  {author} {\bibinfo {author} {\bibfnamefont {W.~Gavin}\
  \bibnamefont {Williams}},\ }\href@noop {} {\emph {\bibinfo {title} {Polarized
  Neutrons}}}\ (\bibinfo  {publisher} {Clarendon , Oxford},\ \bibinfo {year}
  {1988})\BibitemShut {NoStop}%
\bibitem [{\citenamefont {Lovesey}(1984)}]{Lovesey_book}%
  \BibitemOpen
  \bibfield  {author} {\bibinfo {author} {\bibfnamefont {Stephen~William}\
  \bibnamefont {Lovesey}},\ }\href@noop {} {\emph {\bibinfo {title} {Theory of
  neutron scattering from condensed matter}}}\ (\bibinfo  {publisher}
  {Clarendon},\ \bibinfo {address} {Oxford},\ \bibinfo {year}
  {1984})\BibitemShut {NoStop}%
\bibitem [{\citenamefont {Toth}\ and\ \citenamefont {Lake}(2015)}]{Toth15}%
  \BibitemOpen
  \bibfield  {author} {\bibinfo {author} {\bibfnamefont {S}~\bibnamefont
  {Toth}}\ and\ \bibinfo {author} {\bibfnamefont {B}~\bibnamefont {Lake}},\
  }\bibfield  {title} {\enquote {\bibinfo {title} {Linear spin wave theory for
  single-q incommensurate magnetic structures},}\ }\href
  {http://stacks.iop.org/0953-8984/27/i=16/a=166002} {\bibfield  {journal}
  {\bibinfo  {journal} {J. Phys.: Cond. Matt.}\ }\textbf {\bibinfo {volume}
  {27}},\ \bibinfo {pages} {166002} (\bibinfo {year} {2015})}\BibitemShut
  {NoStop}%
\bibitem [{\citenamefont {Kreisel}\ \emph {et~al.}(2008)\citenamefont
  {Kreisel}, \citenamefont {Sauli}, \citenamefont {Hasselmann},\ and\
  \citenamefont {Kopietz}}]{Kreisel2008}%
  \BibitemOpen
  \bibfield  {author} {\bibinfo {author} {\bibfnamefont {Andreas}\ \bibnamefont
  {Kreisel}}, \bibinfo {author} {\bibfnamefont {Francesca}\ \bibnamefont
  {Sauli}}, \bibinfo {author} {\bibfnamefont {Nils}\ \bibnamefont
  {Hasselmann}}, \ and\ \bibinfo {author} {\bibfnamefont {Peter}\ \bibnamefont
  {Kopietz}},\ }\bibfield  {title} {\enquote {\bibinfo {title} {Quantum
  heisenberg antiferromagnets in a uniform magnetic field: Nonanalytic magnetic
  field dependence of the magnon spectrum},}\ }\href {\doibase
  10.1103/PhysRevB.78.035127} {\bibfield  {journal} {\bibinfo  {journal} {Phys.
  Rev. B}\ }\textbf {\bibinfo {volume} {78}},\ \bibinfo {pages} {035127}
  (\bibinfo {year} {2008})}\BibitemShut {NoStop}%
\bibitem [{\citenamefont {Laurence}\ and\ \citenamefont
  {Petitgrand}(1973)}]{Laurence73}%
  \BibitemOpen
  \bibfield  {author} {\bibinfo {author} {\bibfnamefont {G.}~\bibnamefont
  {Laurence}}\ and\ \bibinfo {author} {\bibfnamefont {D.}~\bibnamefont
  {Petitgrand}},\ }\bibfield  {title} {\enquote {\bibinfo {title} {Thermal
  conductivity and magnon-phonon resonant interaction in antiferromagnetic
  {FeCl}$_{2}$},}\ }\href {\doibase 10.1103/PhysRevB.8.2130} {\bibfield
  {journal} {\bibinfo  {journal} {Phys. Rev. B}\ }\textbf {\bibinfo {volume}
  {8}},\ \bibinfo {pages} {2130--2138} (\bibinfo {year} {1973})}\BibitemShut
  {NoStop}%
\bibitem [{\citenamefont {Boiteux}\ \emph {et~al.}(1972)\citenamefont
  {Boiteux}, \citenamefont {Doussineau}, \citenamefont {Ferry},\ and\
  \citenamefont {H\"ochli}}]{Boiteux72}%
  \BibitemOpen
  \bibfield  {author} {\bibinfo {author} {\bibfnamefont {M.}~\bibnamefont
  {Boiteux}}, \bibinfo {author} {\bibfnamefont {P.}~\bibnamefont {Doussineau}},
  \bibinfo {author} {\bibfnamefont {B.}~\bibnamefont {Ferry}}, \ and\ \bibinfo
  {author} {\bibfnamefont {U.~T.}\ \bibnamefont {H\"ochli}},\ }\bibfield
  {title} {\enquote {\bibinfo {title} {Antiferroacoustic resonances and
  magnetoelastic coupling in {GdAlO}$_{3}$},}\ }\href {\doibase
  10.1103/PhysRevB.6.2752} {\bibfield  {journal} {\bibinfo  {journal} {Phys.
  Rev. B}\ }\textbf {\bibinfo {volume} {6}},\ \bibinfo {pages} {2752--2762}
  (\bibinfo {year} {1972})}\BibitemShut {NoStop}%
\bibitem [{\citenamefont {Callen}\ and\ \citenamefont
  {Callen}(1965)}]{Callen65}%
  \BibitemOpen
  \bibfield  {author} {\bibinfo {author} {\bibfnamefont {Earl}\ \bibnamefont
  {Callen}}\ and\ \bibinfo {author} {\bibfnamefont {Herbert~B.}\ \bibnamefont
  {Callen}},\ }\bibfield  {title} {\enquote {\bibinfo {title}
  {Magnetostriction, forced magnetostriction, and anomalous thermal expansion
  in ferromagnets},}\ }\href {\doibase 10.1103/PhysRev.139.A455} {\bibfield
  {journal} {\bibinfo  {journal} {Phys. Rev.}\ }\textbf {\bibinfo {volume}
  {139}},\ \bibinfo {pages} {A455--A471} (\bibinfo {year} {1965})}\BibitemShut
  {NoStop}%
\bibitem [{\citenamefont {Evenson}\ and\ \citenamefont
  {Liu}(1969)}]{Evenson69}%
  \BibitemOpen
  \bibfield  {author} {\bibinfo {author} {\bibfnamefont {W.~E.}\ \bibnamefont
  {Evenson}}\ and\ \bibinfo {author} {\bibfnamefont {S.~H.}\ \bibnamefont
  {Liu}},\ }\bibfield  {title} {\enquote {\bibinfo {title} {Theory of magnetic
  ordering in the heavy rare earths},}\ }\href {\doibase
  10.1103/PhysRev.178.783} {\bibfield  {journal} {\bibinfo  {journal} {Phys.
  Rev.}\ }\textbf {\bibinfo {volume} {178}},\ \bibinfo {pages} {783--794}
  (\bibinfo {year} {1969})}\BibitemShut {NoStop}%
\bibitem [{\citenamefont {Jensen}(1971)}]{Jensen_risoe}%
  \BibitemOpen
  \bibfield  {author} {\bibinfo {author} {\bibfnamefont {Jens}\ \bibnamefont
  {Jensen}},\ }\href
  {http://orbit.dtu.dk/services/downloadRegister/53239467/ris_252.pdf} {\emph
  {\bibinfo {title} {Magneto-elastic interactions in the heavy rare-earth
  metals and the elastic properties of terbium}}}\ (\bibinfo  {publisher}
  {Ris\o \ National Laboratory, Denmark},\ \bibinfo {year} {1971})\BibitemShut
  {NoStop}%
\bibitem [{\citenamefont {Colpa}(1978)}]{Colpa78}%
  \BibitemOpen
  \bibfield  {author} {\bibinfo {author} {\bibfnamefont {J.H.P.}\ \bibnamefont
  {Colpa}},\ }\bibfield  {title} {\enquote {\bibinfo {title} {Diagonalization
  of the quadratic boson {Hamiltonian}},}\ }\href {\doibase
  http://dx.doi.org/10.1016/0378-4371(78)90160-7} {\bibfield  {journal}
  {\bibinfo  {journal} {Physica A}\ }\textbf {\bibinfo {volume} {93}},\
  \bibinfo {pages} {327 -- 353} (\bibinfo {year} {1978})}\BibitemShut {NoStop}%
\bibitem [{\citenamefont {Serga}\ \emph {et~al.}(2012)\citenamefont {Serga},
  \citenamefont {Sandweg}, \citenamefont {Vasyuchka}, \citenamefont
  {Jungfleisch}, \citenamefont {Hillebrands}, \citenamefont {Kreisel},
  \citenamefont {Kopietz},\ and\ \citenamefont {Kostylev}}]{Serga12}%
  \BibitemOpen
  \bibfield  {author} {\bibinfo {author} {\bibfnamefont {A.~A.}\ \bibnamefont
  {Serga}}, \bibinfo {author} {\bibfnamefont {C.~W.}\ \bibnamefont {Sandweg}},
  \bibinfo {author} {\bibfnamefont {V.~I.}\ \bibnamefont {Vasyuchka}}, \bibinfo
  {author} {\bibfnamefont {M.~B.}\ \bibnamefont {Jungfleisch}}, \bibinfo
  {author} {\bibfnamefont {B.}~\bibnamefont {Hillebrands}}, \bibinfo {author}
  {\bibfnamefont {A.}~\bibnamefont {Kreisel}}, \bibinfo {author} {\bibfnamefont
  {P.}~\bibnamefont {Kopietz}}, \ and\ \bibinfo {author} {\bibfnamefont
  {M.~P.}\ \bibnamefont {Kostylev}},\ }\bibfield  {title} {\enquote {\bibinfo
  {title} {Brillouin light scattering spectroscopy of parametrically excited
  dipole-exchange magnons},}\ }\href {\doibase 10.1103/PhysRevB.86.134403}
  {\bibfield  {journal} {\bibinfo  {journal} {Phys. Rev. B}\ }\textbf {\bibinfo
  {volume} {86}},\ \bibinfo {pages} {134403} (\bibinfo {year}
  {2012})}\BibitemShut {NoStop}%
\bibitem [{\citenamefont {Toulouse}\ \emph {et~al.}(2014)\citenamefont
  {Toulouse}, \citenamefont {Liu}, \citenamefont {Gallais}, \citenamefont
  {Measson}, \citenamefont {Sacuto}, \citenamefont {Cazayous}, \citenamefont
  {Chaix}, \citenamefont {Simonet}, \citenamefont {de~Brion}, \citenamefont
  {Pinsard-Godart}, \citenamefont {Willaert}, \citenamefont {Brubach},
  \citenamefont {Roy},\ and\ \citenamefont {Petit}}]{Toulouse2014}%
  \BibitemOpen
  \bibfield  {author} {\bibinfo {author} {\bibfnamefont {C.}~\bibnamefont
  {Toulouse}}, \bibinfo {author} {\bibfnamefont {J.}~\bibnamefont {Liu}},
  \bibinfo {author} {\bibfnamefont {Y.}~\bibnamefont {Gallais}}, \bibinfo
  {author} {\bibfnamefont {M.-A.}\ \bibnamefont {Measson}}, \bibinfo {author}
  {\bibfnamefont {A.}~\bibnamefont {Sacuto}}, \bibinfo {author} {\bibfnamefont
  {M.}~\bibnamefont {Cazayous}}, \bibinfo {author} {\bibfnamefont
  {L.}~\bibnamefont {Chaix}}, \bibinfo {author} {\bibfnamefont
  {V.}~\bibnamefont {Simonet}}, \bibinfo {author} {\bibfnamefont
  {S.}~\bibnamefont {de~Brion}}, \bibinfo {author} {\bibfnamefont
  {L.}~\bibnamefont {Pinsard-Godart}}, \bibinfo {author} {\bibfnamefont
  {F.}~\bibnamefont {Willaert}}, \bibinfo {author} {\bibfnamefont {J.~B.}\
  \bibnamefont {Brubach}}, \bibinfo {author} {\bibfnamefont {P.}~\bibnamefont
  {Roy}}, \ and\ \bibinfo {author} {\bibfnamefont {S.}~\bibnamefont {Petit}},\
  }\bibfield  {title} {\enquote {\bibinfo {title} {Lattice and spin excitations
  in multiferroic $h$-{YMnO}$_{3}$},}\ }\href {\doibase
  10.1103/PhysRevB.89.094415} {\bibfield  {journal} {\bibinfo  {journal} {Phys.
  Rev. B}\ }\textbf {\bibinfo {volume} {89}},\ \bibinfo {pages} {094415}
  (\bibinfo {year} {2014})}\BibitemShut {NoStop}%
\bibitem [{\citenamefont {Fabr\`eges}(2010)}]{Fabreges_PhD}%
  \BibitemOpen
  \bibfield  {author} {\bibinfo {author} {\bibfnamefont {Xavier}\ \bibnamefont
  {Fabr\`eges}},\ }\emph {\bibinfo {title} {Etude des propri\'et\'es
  magn\'etiques et du couplage spin/r\'eseau dans les compos\'es
  multiferro\"iques RMnO$_3$ hexagonaux par diffusion de neutrons}},\ \href
  {https://tel.archives-ouvertes.fr/tel-00551271v4} {Ph.D. thesis},\ \bibinfo
  {school} {Universit\'e Paris Sud - Paris XI} (\bibinfo {year}
  {2010})\BibitemShut {NoStop}%
\bibitem [{\citenamefont {Prince}(2006)}]{TabCrys}%
  \BibitemOpen
  \bibinfo {editor} {\bibfnamefont {E.}~\bibnamefont {Prince}},\ ed.,\ \href
  {\doibase 10.1107/97809553602060000103} {\emph {\bibinfo {title}
  {International Tables for Crystallography, Vol. C}}}\ (\bibinfo  {publisher}
  {Wiley},\ \bibinfo {year} {2006})\BibitemShut {NoStop}%
\bibitem [{\citenamefont {Park}\ \emph {et~al.}(2002)\citenamefont {Park},
  \citenamefont {Kong}, \citenamefont {Pirogov}, \citenamefont {Choi},
  \citenamefont {Park}, \citenamefont {Choi}, \citenamefont {Lee},\ and\
  \citenamefont {Jo}}]{Park2002}%
  \BibitemOpen
  \bibfield  {author} {\bibinfo {author} {\bibfnamefont {J.}~\bibnamefont
  {Park}}, \bibinfo {author} {\bibfnamefont {U.}~\bibnamefont {Kong}}, \bibinfo
  {author} {\bibfnamefont {A.}~\bibnamefont {Pirogov}}, \bibinfo {author}
  {\bibfnamefont {S.I.}\ \bibnamefont {Choi}}, \bibinfo {author} {\bibfnamefont
  {J.-G.}\ \bibnamefont {Park}}, \bibinfo {author} {\bibfnamefont {Y.N.}\
  \bibnamefont {Choi}}, \bibinfo {author} {\bibfnamefont {C.}~\bibnamefont
  {Lee}}, \ and\ \bibinfo {author} {\bibfnamefont {W.}~\bibnamefont {Jo}},\
  }\bibfield  {title} {\enquote {\bibinfo {title} {Neutron-diffraction studies
  of {YMnO$_3$}},}\ }\href {\doibase 10.1007/s003390201806} {\bibfield
  {journal} {\bibinfo  {journal} {Applied Physics A}\ }\textbf {\bibinfo
  {volume} {74}},\ \bibinfo {pages} {s796--s798} (\bibinfo {year}
  {2002})}\BibitemShut {NoStop}%
\bibitem [{\citenamefont {Fiebig}\ \emph {et~al.}(2000)\citenamefont {Fiebig},
  \citenamefont {Fr\"ohlich}, \citenamefont {Kohn}, \citenamefont {Leute},
  \citenamefont {Lottermoser}, \citenamefont {Pavlov},\ and\ \citenamefont
  {Pisarev}}]{Fiebig2000}%
  \BibitemOpen
  \bibfield  {author} {\bibinfo {author} {\bibfnamefont {M.}~\bibnamefont
  {Fiebig}}, \bibinfo {author} {\bibfnamefont {D.}~\bibnamefont {Fr\"ohlich}},
  \bibinfo {author} {\bibfnamefont {K.}~\bibnamefont {Kohn}}, \bibinfo {author}
  {\bibfnamefont {St.}\ \bibnamefont {Leute}}, \bibinfo {author} {\bibfnamefont
  {Th.}\ \bibnamefont {Lottermoser}}, \bibinfo {author} {\bibfnamefont {V.~V.}\
  \bibnamefont {Pavlov}}, \ and\ \bibinfo {author} {\bibfnamefont {R.~V.}\
  \bibnamefont {Pisarev}},\ }\bibfield  {title} {\enquote {\bibinfo {title}
  {Determination of the magnetic symmetry of hexagonal manganites by second
  harmonic generation},}\ }\href {\doibase 10.1103/PhysRevLett.84.5620}
  {\bibfield  {journal} {\bibinfo  {journal} {Phys. Rev. Lett.}\ }\textbf
  {\bibinfo {volume} {84}},\ \bibinfo {pages} {5620--5623} (\bibinfo {year}
  {2000})}\BibitemShut {NoStop}%
\bibitem [{\citenamefont {Martinsson}\ and\ \citenamefont
  {Movchan}(2003)}]{Martinsson03}%
  \BibitemOpen
  \bibfield  {author} {\bibinfo {author} {\bibfnamefont {P.~G.}\ \bibnamefont
  {Martinsson}}\ and\ \bibinfo {author} {\bibfnamefont {A.~B.}\ \bibnamefont
  {Movchan}},\ }\bibfield  {title} {\enquote {\bibinfo {title} {Vibrations of
  lattice structures and phononic band gaps},}\ }\href {\doibase
  10.1093/qjmam/56.1.45} {\bibfield  {journal} {\bibinfo  {journal} {Quarterly
  J. Mech. Appl. Math.}\ }\textbf {\bibinfo {volume} {56}},\ \bibinfo {pages}
  {45--64} (\bibinfo {year} {2003})}\BibitemShut {NoStop}%
\bibitem [{\citenamefont {Spremo}\ \emph {et~al.}(2005)\citenamefont {Spremo},
  \citenamefont {Sch\"utz}, \citenamefont {Kopietz}, \citenamefont
  {Pashchenko}, \citenamefont {Wolf}, \citenamefont {Lang}, \citenamefont
  {Bats}, \citenamefont {Hu},\ and\ \citenamefont {Schmidt}}]{Spremo2005}%
  \BibitemOpen
  \bibfield  {author} {\bibinfo {author} {\bibfnamefont {Ivan}\ \bibnamefont
  {Spremo}}, \bibinfo {author} {\bibfnamefont {Florian}\ \bibnamefont
  {Sch\"utz}}, \bibinfo {author} {\bibfnamefont {Peter}\ \bibnamefont
  {Kopietz}}, \bibinfo {author} {\bibfnamefont {Volodymyr}\ \bibnamefont
  {Pashchenko}}, \bibinfo {author} {\bibfnamefont {Bernd}\ \bibnamefont
  {Wolf}}, \bibinfo {author} {\bibfnamefont {Michael}\ \bibnamefont {Lang}},
  \bibinfo {author} {\bibfnamefont {Jan~W.}\ \bibnamefont {Bats}}, \bibinfo
  {author} {\bibfnamefont {Chunhua}\ \bibnamefont {Hu}}, \ and\ \bibinfo
  {author} {\bibfnamefont {Martin~U.}\ \bibnamefont {Schmidt}},\ }\bibfield
  {title} {\enquote {\bibinfo {title} {Magnetic properties of a metal-organic
  antiferromagnet on a distorted honeycomb lattice},}\ }\href {\doibase
  10.1103/PhysRevB.72.174429} {\bibfield  {journal} {\bibinfo  {journal} {Phys.
  Rev. B}\ }\textbf {\bibinfo {volume} {72}},\ \bibinfo {pages} {174429}
  (\bibinfo {year} {2005})}\BibitemShut {NoStop}%
\end{thebibliography}
\end{document}